\documentstyle[12pt]{article}
\newcommand{\be}{\begin{equation}}
\newcommand{\ee}{\end{equation}}
\newcommand{\bea}{\begin{eqnarray}}
\newcommand{\eea}{\end{eqnarray}}
\newcommand{\ba}{\begin{array}}
\newcommand{\ea}{\end{array}}
\newcommand{\vier}{\\ [4 pt]}

\newcommand{\halb}{{1\over2}}
\newcommand{\quart}{{1\over4}}
\newcommand{\th}{\theta}

\newcommand{\imag}{{\rm Im}\hspace{2pt}}
\newcommand{\ds}{\displaystyle}
\newcommand{\sss}{\scriptstyle}
\newcommand{\rp}{\IR^+_0}
\newcommand{\hq}{\hspace{5pt}}
\def\inbar{\vrule height1.5ex width.4pt depth0pt}
\def\IB{\relax{\rm I\kern-.18em B}}
\def\IC{\relax\,\hbox{$\inbar\kern-.3em{\rm C}$}}
\def\ID{\relax{\rm I\kern-.18em D}}
\def\IE{\relax{\rm I\kern-.18em E}}
\def\IF{\relax{\rm I\kern-.18em F}}
\def\IG{\relax\,\hbox{$\inbar\kern-.3em{\rm G}$}}
\def\IH{\relax{\rm I\kern-.18em H}}
\def\II{\relax{\rm I\kern-.18em I}}
\def\IK{\relax{\rm I\kern-.18em K}}
\def\IL{\relax{\rm I\kern-.18em L}}
\def\IM{\relax{\rm I\kern-.18em M}}
\def\IN{\relax{\rm I\kern-.18em N}}
\def\IO{\relax\,\hbox{$\inbar\kern-.3em{\rm O}$}}
\def\IP{\relax{\rm I\kern-.18em P}}
\def\IQ{\relax\,\hbox{$\inbar\kern-.3em{\rm Q}$}}
\def\IR{\relax{\rm I\kern-.18em R}}
\def\ZZ{\relax{\sf Z\kern-.4em Z}}

\begin{document}

\title{Comment on the Generation Number in Orbifold
       Compactifications\thanks{Supported by Deutsche Forschungsgemeinschaft}}

\author{{\sc J. Erler} and {\sc A. Klemm} \\ \\
       {\em Physik Department} \\
       {\em Technische Universit\"at M\"unchen} \\
       {\em 8046 Garching, Germany} \\ }

\maketitle

\begin{picture}(5,2.5)(-320,-420)
\put(12,-90){MPI--Ph/92-60}
\put(12,-105){TUM--TH--146/92}
\put(12,-123){June 1992}
\end{picture}

\begin{abstract}
There has been some confusion concerning the number of $(1,1)$-forms
in orbifold compactifications of the heterotic string in numerous
publications. In this note we point out the relevance of the underlying
torus lattice on this number. We answer the question when different lattices
mimic the same physics and when this is not the case. As a byproduct we
classify all symmetric $Z_N$-orbifolds with $(2,2)$ world sheet supersymmetry
obtaining also some new ones.
\end{abstract}

\section{Introduction}

String compactifications on toroidal $Z_N$-orbifolds~\cite{DHVW}
are among the most intensively studied ones. They provide us with the simplest
string models  which have semi realistic features.
Because the one loop partition function and the couplings can be
calculated explicitly in dependence of the untwisted moduli,
many generic properties concerning the string moduli space and the
effective low energy theory can be investigated here in
detail\footnote{The situation for more general compactification
schemes has improved, as P.~Candelas, X.~De la Ossa, P.~Green and
L.~Parkes worked out the modulus dependence explicetly~\cite{COGP} for the
quintic
threefold in $\IP^4$. Other Calabi-Yau manifolds
were investigated in~\cite{CY}.}.
Including all background parameters in the framework of heterotic
compactifications and allowing for the most general twists, a rich class
of models, with partly phenomenological very attractive features,
emerges. The question is still open, whether some  standard  string model,
which can be related in a painless manner to the known phenomenology,
is contained in this class.

Toroidal orbifolds have also attracted the attention of the
mathematicians, because the partition functions of $(2,2)$ models
contain informations, which can be interpreted as topological
data of a Calabi-Yau manifold. The latter can indeed be
constructed by a certain resolving process of the
orbi\-fold singularities, which establishes an exciting relation
between singularity theory and the theory of modular functions.

In this situation it comes as a surprise that some of the most
fundamental properties from the physical as well as from the
mathematical point of view, namely the individual numbers of generations and
antigenerations, i.e. $(1,1)$-forms and $(2,1)$-forms
respectively, on the orbifold have been reported incorrectly
in the literature.

$Z_N$-orbifolds~\cite{DHVW} are unambiguously defined through a twist acting as
an automorphism in some torus lattice. Clearly, a given twist matrix determines
both the spectrum of twist eigenvalues and the lattices possessing the
automorphism. Some properties like the number of chiral generations, the number
of space time supersymmetries or the number of untwisted moduli fields
only depend on these eigenvalues. On the other hand, however, the role of the
underlying lattice has been underestimated in the past. For instance,
it determines the modular symmetry group~\cite{GRV,SW,LMN,EJN,Munich},
which has attracted much attention due to its importance for discussions of
low energy effective actions~\cite{Ferrara}.

In this communication we will point out that the lattice has even impact
on the number of $(1,1)$-forms ($h_{1,1}$) and thus on the massless spectrum.
Since this fact has been overlooked up to now, the true total number of 27's
and ${\overline {27}}$'s depart from those stated in the literature. Especially
the Lie-algebra lattices assigned to a stated $h_{1,1}$ differ from the results
of this publication. Neglecting the dependence of orbifold properties on
the lattice, has also led to an incomplete classification of symmetric
$Z_N$-orbifolds with $(2,2)$ world sheet supersymmetry and vanishing
discrete background field.
We have found $18$ inequivalent orbifolds in this class\footnote{There
exists a more natural and much more efficient way of
constructing $(2,2)$ string compactifications using the
Landau-Ginzburg approach. A classification of these string vacua was
performed in \cite{KSKS}.}.

One should stress, that the correct treatment does not lead to any changes
for the two prime orbifolds $Z_3$ and $Z_7$, where no fixed tori occur.
In the case of non prime orbifolds we have typically the situation, that
there exits a lattice to which apply parts of the reasoning that appear in
the literature. This lattice is, however, usually not the
one the authors refer to.

There is a variety of methods to obtain informations about the spectra
of $Z_N$-orbifolds.
One is to study the possible resolutions of orbifold
singularities~\cite{MOP,RY,A}. In three (complex) dimensions
these are either related to fixed points or to fixed curves. The cases of
fixed points are completely understood. In the presence of fixed curves,
however, more care is needed and different torus lattices give rise
to different resolutions.

Another possibility is to construct
the one loop partition function as done in~\cite{IMNQ}. This is equivalent
to knowing all
massless and massive states\footnote{In more general constructions
like $Z_N \times Z_M$ or non-abelian  orbifolds one-loop modular
invariance might be insufficient for all loop modular invariance
and also for defining the model completely~\cite{Vafa}.}.
We will show that those parts of
the partition function which are related to sectors, where the corresponding
twist matrix leaves fixed tori, have a somewhat different structure.

Finally, one can construct twist invariant vertex operators by using the
mode expansions
of the untwisted and twisted coordinates~\cite{KKKOOT}. Again we will argue
that in sectors with fixed directions different solutions arise whenever
the underlying lattice is changed.

Since the main purpose of this note is to emphasize the importance of the
compactification lattice, we will construct the complete list of
symmetric $Z_N$-orbifolds of $(2,2)$-type with vanishing discrete background
fields $B_{\mu\nu}$. As shown in~\cite{EJN} non-vanishing discrete
background fields very much mimic asymmetric orbifolds. It was also shown
there that they can sometimes be transformed to orbifolds without
such backgrounds. If this is the case the defining torus lattice in
the transformed model typically differs from the original one. Thus we are
also led to discuss these class of models.

The organization of the paper is as follows: In section~\ref{class} we
construct in a systematic way all models of the above mentioned type. This
includes all Coxeter-twists but also more general ones. Using methods
developed in~\cite{EJN}, we will also recognize equivalences of models
considered as different before~\cite{KKKOOT}. Section~\ref{orbi} is devoted
to the proof that $h_{1,1}$ depends on the chosen lattice. It uses the
one loop partition function as described above and discusses also some
aspects of the mode expansion approach. In section~\ref{geometry}
we confirm our results by a completely different method utilizing
results of singularity theory. We give an easy prescription
how to calculate $h_{1,1}$ and $h_{1,2}$ from the fixed sets.

\section{Classification of the $Z_N-$Orbifolds}
\label{class}

In this section we classify $(2,2)$ string theories on orbifolds,
which can be obtained by dividing out a $Z_N$ group in a symmetric
way from a six dimensional torus $T^6$ with vanishing discrete $B$-field.

To be more precise we will classify equivalence classes defined up
to modular deformations. We start the discussion with a short
extraction of the necessary concepts for toroidal orbifolds, mainly from the
physics literature.

\subsection{Concepts in toroidal- and orbifold compactifications}
Let $\Lambda =\{\sum_{i=1}^6 n^i e_i|n_i\in \ZZ\}$ be a lattice embedded
in Euclidean space $\IR^6$. Due to the canonical isomorphism we denote
the basis of the tangent space also by $e_i$  and define the induced
metric as $g_{i\, j}=\langle e_i, e_j\rangle$, where
$\langle\, ,\, \rangle$ is the Euclidean scalar product. We denote the
dual lattice w.r.t.\ $\langle\, ,\, \rangle$ by $\Lambda^*$.
The six torus is defined as the quotient of
$\IR^6$ w.r.t.\ $\Lambda$,
\be
T^6:=\IR^6/\Lambda\label{ref}.
\ee
One may also introduce an antisymmetric background field $B=b_{i\,j}
{e}^i\wedge {e}^j$ as a geometrical data of the torus.
The allowed momenta for the left and right
movers of strings compactified on the torus with this background field
are given by~\cite{Narain}
\be
P_L={1\over 2} m + g n - bn, \qquad P_R={1\over 2} m - g n-bn\label{prpl},
 \ee
where $n$ and $m$ are integer six vectors describing the winding
and momentum quantum numbers of the string state. They label the elements of
$\Lambda$ and its dual $\Lambda^*$ respectively.
The geometry of the underlying torus, i.e. the so called {\sl modular
parameters}\footnote{In the heterotic theory
other modular parameters appear in form of Wilson lines $A_i^I$.}
$g_{i\, j}$ and $b_{i\, j}$, enters string theory only via
bilinear forms of $(m,n)$
describing the scaling dimension,
\be \ba{rl}
H_{(g,b)}(m,n;m,n)&=P_L g^{-1} P_L+P_R g^{-1} P_R\\[ 3 mm]
&= \displaystyle{
{1\over 2} m^T g^{-1}m+ 2 n^Tg\, n-
m^T\, g^{-1} b\, n+
n^T\, b g^{-1}\, m +
2 n^T b^{-1} g\, n}
\ea
\label{ener}
\ee
and the spin,
\be
S(m,n;m,n)=P_L g^{-1} P_L-P_R g^{-1} P_R=2\, m^T n,\label{spin}
\ee
of the physical vertex operators, up to geometry independent
oscillator contributions.
A linear relabeling of $(n,m)$ leaving
invariant the bilinear form (\ref{spin}) can be accompanied by a
relabeling of the the modular parameters $g$ and  $b$  such that
also the scaling dimensions (\ref{ener})
of the vertex operators are invariant.
That is, string theory is the same for geometrically different tori and this
leads us to following definition:

\newtheorem{definition}{Definition}
\begin{definition}
The bijective linear integer transformations of $(n,m)\mapsto
gl(n,m)$, which leave (\ref{spin}) invariant,
generate an invariance group $G$ for string theory on the torus.

The requirement that $H$ stays invariant
$H_{(g,b)} (n,m;n,m)=H_{(g',b')}(gl(n,m);$ $gl(n,m))$ induces
a representation of $G$ as transformations of the moduli
$\rho(gl):(g,b)\mapsto (g', b')$, which is
not\footnote{Obviously $gl(n,m)=(-n,-m)$ is in the kernel of $\rho$.}
faithful.
The group generated by this transformations is called modular symmetry
group $\cal G$.
\end{definition}
Setting the $B$-field to zero and restricting
to a subgroup of $G$, which do not mix windings and
momenta~\cite{GRV} we have the following
\newtheorem{symequiv}{Lemma}
\begin{symequiv}
Let $M\in GL(6,\ZZ)$. The
string theory on $T^6$ remains invariant under the simultaneous
transformations $n\rightarrow M\, n$, $m\rightarrow {M^T}^{-1}\, m$ and
$g \rightarrow {M^T}^{-1}\, g\, M^{-1}$.
\end{symequiv}

An orbifold is defined by a finite group $T$ generated by elements
$\Theta_i$ of $G$, all of which leave $H_{(g_o,b_o)}$
($(g_o,b_o)\neq \emptyset $) invariant.
The latter requirement defines $(g_o,b_o)$ the untwisted moduli space of the
orbifold, which is a subspace of the moduli space $(g,b)$ of the torus.
$(g_o,b_o)$ might be a point in moduli space,
but in general the $\Theta_i$ specify the orbifold
only up to modular deformations.

The symmetric $Z_N$ orbifolds with vanishing discrete $B$-field, which we
want to classify amount to the simplest possible choice for $\Theta$.
Namely we pick an element $\theta\in GL(6,\ZZ)$ with $\theta^N=1$ and
transform simultaneously\footnote{Note that these symmetries
are directly realised as discrete automorphism on the target-space torus,
whereas symmetries which mix momenta and windings are specific for
string theory. It might be interesting for
geometers to consider also the latter ones,
as orbifoldizing w.r.t.\ some of them also give
rise to Calabi-Yau manifolds. Moreover the symmetries which lead to the
mirrors of the $Z_N$-orbifolds are also in this class.}
$n\rightarrow\theta n$ and $m\rightarrow{\theta^T}^{-1} m$.
The spin $S$ is invariant
by construction and requiring invariance of
$H_{g_o}$ we get
\be
\theta^T g_o\, \theta=g_o.\label{cris}
\ee
Condition (\ref{cris}) ensures that the lattice
automorphism acts cristallographically.

The basis transformation on $(n,m)$ of Lemma 1
accompanied with a transformation of $g\mapsto {M^T}^{-1}\, g\, M^{-1}$
does not change the string theory on the torus and if we
transform a given twist covariantly we also define the same orbifold thereof.
We have therefore the following
equivalence relation for the twist matrices~\cite{EJN}:

\begin{definition} Two twist matrices
$\theta$ and $\theta'$ are equivalent in the sense that
they define the same orbifold string theory, if and only if
$\exists\, M\in GL(6,\ZZ)$ such that $\theta=M\theta' M^{-1}$.
\end{definition}
The modular group of the orbifold can be described as follows\cite{GRV,Munich}
\newtheorem{orbmod}{Corollary}
\begin{orbmod} The $M\in GL(6,\ZZ)$ which fulfill
$\theta^n=M\theta M^{-1}$ ($n\in \ZZ_+$) generate a symmetry group of the
orbifold, which induces a non\footnote{By
(\ref{cris}) all powers of $\theta$ are in the kernel of $\rho$.}
faithful representation
on the orbifold moduli $\rho(M):g_o\mapsto g_o'$ by  the requirement
$H_{g_o}(n,m;n,m) = H_{g_o'}(M n,{M^T}^{-1} m; M n,{M^T}^{-1} m)$.
\end{orbmod}

\subsection{Description of the Classification}
We start the classification of the symmetric $Z_N$ twists by
analysing their possible eigenvalues.
Let $\theta$ be a $Z_N$ twist in an
n-dimensional lattice. By a transformation
matrix $B\in GL(n,\IC)$ we can pass
from the lattice basis where the twist $\theta$
is integer valued
to a basis where it is diagonal
\be
\theta_d=B^{-1}\theta B={\rm diag}(\xi^{a_1},\ldots,\xi^{a_n}).\label{diag}
\ee
Here $\xi=e^{2 \pi i\over N}$ is the N'th order root of unity and
$a_i\in \IN_0$, $a_i<N$.
Since we wish $\theta$ to act as an integer matrix in some lattice we
get as a necessary condition on the exponents $\lambda_i=a_i/N$
of $\theta_d$ that they define a characteristic polynomial $P(x)$
{\em over the integers\/}.
Especially the {\sl Lefschetz Fixed Point Theorem\/}
which gives the number of fixed points\footnote{${\rm det}(1-\theta)=0$
in (\ref{lft}) signals the occurrence of fixed sublattices.}
$\chi_F$ of a lattice automorphism $\theta$ as
\be
\chi_F={\rm det}(1-\theta),\label{lft}
\ee
implies $P(1)\in \ZZ$.

First we search for $\theta_d$, which fulfill the necessary condition
above. Let us denote by $(a,b)$ the greatest common divisor of $a$ and $b$.
We call $\theta_d$ {\em proper\/}, if $(a_i,N)=1$ for all $i$.
Obviously all $\theta_d$ can be constructed from {\em proper}
subblocks $\theta'_d$.
The Euler function is defined $\phi(N)$ as
the number of integers $0<l<N$ with $(l,n)$=1.
The characteristic polynomial for a {\em proper\/} twist
is a so called cyclotomic polynomial of degree $\phi(N)$
\cite{HMV}
\be
P_N(x)=\prod_{{{0<a<N}\atop {(a,N)=1}}} (x-\xi^a),
\ee
which  enjoys the following properties
\begin{itemize}
\begin{enumerate}
\item $P_N(x)$ is in the polynom ring over the integers: $P_N(x)\in \ZZ[x]$.
\item $P_N(x)$ is irreducible in $\ZZ[x]$.
\item $P_N(1)=\cases{p& if $N=p^r$ with $p$ prime,\cr 1& otherwise.}$
\end{enumerate}
\end{itemize}
1.\ ensures our necessary condition for $\theta'$ to act
as a lattice automorphism.
Due to  2.\ the minimal dimension in which
a $Z_N$ twist can be realised cristallographically
and {\em proper\/} is given by $\phi(N)$.
Finally, 3.\ provides us with an effective method for computing
the number of fixed points.
%
%
The dimensions
of the {\em proper} blocks are given by
\be
\phi(N)=N\prod_{i} {{p_i-1}\over p_i}, \quad {\rm where}\,\, p_i {\rm \,\,
are\,\, the\,\,
distinct\,\, prime\,\, factors\,\, of\,\,} N,\label{phin}
\ee
which reduces to $N-1$ for prime $N$.
{}From that we see that for $d=1$ only $Z_2$ with $2$ fixed points
is allowed and furthermore that no other block can be defined
{\em properly} in odd dimensions.
We can easily  construct a table of exponents up to $d=6$.
Because all $\xi^{a_i}$ appear together
with $\overline{\xi^{a_i}}$, we only list half of them. Using (\ref{phin})
it is easy to see that no higher orders are possible.

\vskip  .3 cm
\centerline{Table 1: Exponents of proper $Z_N$ twists}
\vskip  .3 cm
\begin{center}
\begin{tabular}{|rr|rr|rr|}
\hline
\multicolumn{2}{|c|}{$d=2$}
&\multicolumn{2}{|c|}{$d=4$}
&\multicolumn{2}{|c|}{$d=6$}
\\[.31 mm]
\hline
\multicolumn{1}{|c|}{twist}
&\multicolumn{1}{|c|}{$\chi_F$}
&\multicolumn{1}{|c|}{twist}
&\multicolumn{1}{|c|}{$\chi_F$}
&\multicolumn{1}{|c|}{twist}
&\multicolumn{1}{|c|}{$\chi_F$}
 \\[.31 mm]
\hline
\hline &&&&&\\[-3.5 mm]
${1\over 3}(1)$ & 3&       ${1\over 5}(1,2)$ & 5&     ${1\over 7}(1,2,3)$ & 7
\\[ 2 mm]
${1\over 4}(1)$ & 2&       ${1\over 8}(1,3)$ & 2&     ${1\over 9}(1,2,4)$ & 3
\\[ 2 mm]
${1\over 6}(1)$ & 1&       ${1\over 10}(1,3)$ & 1&    ${1\over 14}(1,3,5)$ & 1
\\[ 2 mm]
              &  &         ${1\over 12}(1,5)$  & 1&   ${1\over 18}(1,5,7)$ &
 1\\[ 2 mm]
\hline
\end{tabular}
\end{center}
The allowed sets of exponents in dimensions less then $6$ can now be
obtained by building all possible combinations of the proper ones.
{}From the table we see e.g. that cristallographic automorphisms in
$d=6$  can only exist for
$N=2,3,4,5,6,7,8,9,10,12,14,15,18,20,24,30$. Indeed this list
agrees with the one given  e.g.\ in~\cite{DHVW}.

All eigenvalues appear with their complex conjugate so that
we can define a complex coordinate system
\be
z_i={1\over \sqrt{2}}( x_i+ i x_{i+1}),\qquad i=1,2,3
\label{co-ord}
\ee
on which $\theta$ acts holomorphically.
The condition for obtaining a supersymmetric orbifold can
be formulated in different ways. We follow the geometric approach
\cite{MOP} and require invariance of the holomorphic $(3,0)$ form of the
complex torus (\cite{GH} II.6)
$$\omega =d z_1\wedge d z_2\wedge d z_3,$$
which implies $a_1+a_2+a_3=0\,\, {\rm mod}\,\, N$.
If we furthermore restrict our attention to models which have exactly
$N=1$-supersymmetry, i.e.\ no fixtorus in the first twisted sector, we
get the following $9$ sets of allowed exponents:

\vskip  .3 cm
\centerline{Table 2: Eigenvalues of $Z_N$ twists leaving $N=1$ supersymmetry.}
\vskip  .2 cm
\begin{center}
\begin{tabular}{|rr|rr|rr|}
\hline
\multicolumn{1}{|c|}{twist}
&\multicolumn{1}{|c|}{$\chi_F$}
&\multicolumn{1}{|c|}{twist}
&\multicolumn{1}{|c|}{$\chi_F$}
&\multicolumn{1}{|c|}{twist}
&\multicolumn{1}{|c|}{$\chi_F$}
 \\[.31 mm]
\hline
\hline &&&&&\\[-3.5 mm]
$Z_3:\,\,{1\over 3}(1,1,-2)$  & 27 &
$Z_6':\,\,{1\over 6}(1,2,-3)$ & 12 &
$Z_8':\,\,{1\over 8}(1,3,-4)$ & 8  \\[2 mm]

$Z_4:\,\,{1\over 4}(1,1,-2)$     & 16&
$Z_7:\,\,{1\over 7}(1,2,-3)$     & 7 &
$Z_{12}:\,\,{1\over 12}(1,-5,4)$ & 3 \\[ 2 mm]

$Z_6:\,\,{1\over 6}(1,1,-2)$& 3&
$Z_8:\,\,{1\over 8}(1,-3,2)$& 4&
$Z_{12}':\,\,{1\over 12}(1,5,-6)$& 4  \\ [2 mm]
\hline
\end{tabular}
\end{center}

One approach to find $Z_N$ orbifolds,
which was carried  out in \cite{MOP}, is to
start with a lattice and consider the compatible $Z_N$ autormorphisms.
In contrast we will start with the twist matrix and then specify
the lattice metric as solution to (\ref{cris}).
In virtue of the equivalence relation Definition 2 the action of
an irreducible\footnote{In the following irreducibility is to be
understood over $\ZZ$.}
block of a $Z_N$ twist can be brought into the canonical
form\footnote{In this form $\chi_F$ is simply given by $1-\sum_{i} v_i$.}
\be
\theta=\pmatrix{
               0&&&\ldots&0&v_1\cr
               1&0&&\ldots&0&v_2\cr
               0&1&0&\ldots&0&v_3\cr
               \vdots&&\ddots&&&\vdots\cr
               0&&\ldots&&1&v_n}\label{matf},
\ee
where $v_1=\pm 1$ such that $|{\rm det} (\theta)|=1$. Note
that the form of (\ref{matf}) alone does not imply that the block
is not further reducible.
We calculate the $N$th power of (\ref{matf})
according to our list of possible orders in the given dimension
and search for integer $\vec v$ such that this is the unit matrix.
The corresponding twist matrices can easily be found by
means of a computer program. It turns out that $|v_i|\leq 3$.
In the following table we list only the vectors $\vec v$,
which specify irreducible twist matrices relevant for the
$N=1$ supersymmetric
$Z_N$-orbifolds\footnote{It also turned
out that the cases with $v_1=+1$ exactly correspond to fixed tori.}:

\vskip  .3 cm
\centerline{Table 3: Irreducible building blocks for $Z_N$ twists
relevant for $N=1$ supersymmetry.}
\vskip  .3 cm
\begin{center}
\begin{tabular}{|l|l|l|l|}
\hline
\multicolumn{1}{|c|}{$d=1$}
&\multicolumn{1}{|c|}{$d=2$}
&\multicolumn{1}{|c|}{$d=3$}
&\multicolumn{1}{|c|}{$d=4$}
 \\[.31 mm]
\hline
\hline &&&\\[-3.5 mm]
$Z^{(1)}_2:\,\,(-1)$&
$Z^{(2)}_3:\,\,(-1,-1)$&
$Z^{(3)}_4:\,\,(-1,-1,-1)$&
$Z^{(4)}_3:\,\,(-1,0,-1,0)$
\\[ 2 mm]
&
$Z^{(2)}_4:\,\,(-1,0)$&
$Z^{(3)}_6:\,\,(-1,0,0)$&
$Z^{(4)}_8:\,\,(-1,0,0,0)$
\\[ 2 mm]
&
$Z^{(2)}_6:\,\,(-1,1)$&
&
$Z^{(4)}_{12}:\,\,(-1,0,1,0)$
\\
\hline
\end{tabular}
\vskip  .5 cm
\begin{tabular}{|l|l|}
\hline
\multicolumn{1}{|c|}{$d=5$}
&\multicolumn{1}{|c|}{$d=6$}
 \\[.31 mm]
\hline
\hline &\\[-3.5 mm]
$Z^{(5)}_6:\,\,(-1,-1,-1,-1,-1)$&
$Z_7^{(6)}:\,\,(-1,-1,-1,-1,-1,-1)$
\\[2 mm]
$Z^{(5)}_8:\,\,(-1,-1,0,0,-1)$&
$Z^{(6)}_8:\,\,(-1,0,-1,0,-1,0)$
\\[2 mm]
&
$Z^{(6)}_{12}:\,\,(-1,-1,0,1,0,-1)$
\\[2 mm]
\hline
\end{tabular}
\end{center}

Now we combine these irreducible blocks to $(6\times 6)$ twist matrices
giving rise to $N=1$ supersymmetric orbifolds.

To make contact with the classification of
Coxeter orbifolds in \cite{MOP,IMNQ}, we use the equivalence relation in
Definition 2 to rewrite our twists as Coxeter automorphisms, if possible.
A Weyl reflection\footnote{For the following definitions concerning Lie
algebras \cite{B} is the standard reference.}
is a reflection on the hyperplane perpendicular to a simple root
$$S_i(x)=x-2 {\langle x, e_i\rangle\over  \langle e_i, e_i\rangle} e_i.$$
A Coxeter automorphism  $c$ in a Lie algebra lattice is defined by
sucessive Weyl reflections w.r.t.\ all simple roots $c=S_1\cdot\ldots\cdot
S_{rank}$.
Automorphisms are called outer if they cannot be generated by Weyl reflections.
They are generated by transpositions of roots which are symmetries
of the Dynkin diagram.
Generalised Coxeter automorphisms can be obtained by combining Weyl
reflections with outer automorphisms. We denote a transposition which
exchange the roots $i\leftrightarrow j$ by $P_{ij}$.
In \cite{MOP} generalised Coxeter automorphisms were only considered
if they act in one semisimple factor, e.g.\ in the lattice $A_2\times D_4$
the automorphism $S_1 S_2 S_3 P_{36}P_{35}$ (cyclic permutation of the roots
(3,5,6)).

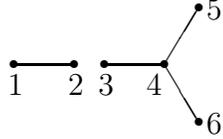
\setlength{\unitlength}{ 4 mm}
\vskip  -1 mm

\begin{picture}(10,4)(-10,0)

\put(2,  1.5){\circle*{0.3}}
\put(1.8,  0.5){1}
\put(4,  1.5){\circle*{0.3}}
\put(3.8,  0.5){2}
\put(2,  1.5){\line(1,0){2}}

\put(5,  1.5){\circle*{0.3}}
\put(4.8,  0.5){3}
\put(7,  1.5){\circle*{0.3}}
\put(6.4,  0.5){4}
\put(5,  1.5){\line(1,0){2}}

\put(8.15,  3.4){\circle*{0.3}}
\put(7,  1.5){\line(3,5){1.2}}
\put(8.4,  3){5}

\put(8.15,  -0.4){\circle*{0.3}}
\put(8.4,   -0.8){6}
\put(7,  1.5){\line(3,-5){1.2}}

\end{picture}
\vskip  7 mm
There is no reason for
this restriction and in fact the full classification involves transpositions
between the semisimple factors, e.g. in $A_3\times A_3$ the automorphism
$S_1S_2S_3P_{16}P_{25}P_{34}$.
As the result\footnote{The relevant combinations of the twist
matrices specified in Table 3 and the
Hodge numbers for the $18$ models can be found in table~\ref{orbtab}.}
of our classification we have the following
\newtheorem{class}{Theorem}
\begin{class}
There exist $18$ inequivalent $N=1$ supersymmetric string theories on
symmetric $Z_N$ orbifolds of $(2,2)$-type without discrete background
all having at least one representative in the class
of generalised Coxeter orbifolds.
More precisely we have $15$ ordinary Co\-xeter orbifolds realized on the
lattices
\begin{center}
\begin{tabular}{|r|r|r|r|}
\hline &&&\\[-3.5 mm]
$A_2\times A_2\times A_2$,&
$A_1\times A_1\times B_2\times B_2$,&
$A_1\times A_3\times B_2$,&
$A_3\times A_3$,\\[ 2 mm]
$A_2\times G_2\times G_2$,&
$A_1\times A_1\times A_2\times G_2$,&
$A_2\times D_4$,&
$A_1\times A_5$,\\[ 2 mm]
$A_6$,&
$B_2\times B_4$,&
$A_1\times A_1\times B_4$,&
$A_1\times D_5$,\\[2 mm]
$A_2\times F_4$,&
$E_6$,&
$D_2\times F_4$.&
\\[2 mm]
\hline
\end{tabular}
\end{center}
and $3$ involving outer automorphisms, namely
$A_1\times A_1\times A_2\times A_2$ with $S_1S_2S_3S_4P_{36}P_{45}$,
$G_2\times A_2\times A_2$ with $S_1S_2S_3S_4P_{36}P_{45}$ and
finally
$A_3\times A_3$ with $S_1S_2S_3P_{16}P_{25}P_{34}$.
\end{class}

This result is to be compared with the result of \cite{MOP}, which
was the basis for further investigations \cite{IMNQ,KKKOOT,CGM}.
Here the authors give in Table 1 of their classification theorem
$13$ examples. They suggest that only $9$ are inequivalent, namely
the one which have different twist eigenvalues (cf table 2 above).
Three pairs of orbifolds which are identified in \cite{MOP},
are inequivalent in the sense of Definition 2. In fact they have
different hodge numbers, as we will explain below.
We agree on the other hand with the identification of the models
$A_2\times D^{[3]}_4$ (with automorphism $S_1 S_2 S_3 P_{36}P_{35}$)
and $A_2\times F_4$ (with Coxeter automorphism) in Table 1 of \cite{MOP}.
Finally we have found {\sl six new\/} examples which are inequivalent to
the ones appearing in \cite{MOP},
three of them as mentioned involve outer
automorphisms between semi simple factors.

\section{Partition Functions}
\label{orbi}
In this section we discuss the construction of one loop partition functions,
which allows for a survey of all string states. For the $E_8 \times E_8$
heterotic string before compactification it is given by~\cite{LT}
\be \label{ZU}
   Z(\tau,\bar{\tau})={1\over 8} {(2 \pi \imag \tau )^{-4} \over
   |\eta(\tau)|^{16}}
   {[ \theta^8_3(\tau)+\theta^8_4(\tau)+\theta^8_2(\tau)]^2 \over
   \eta^{16}(\tau)}
   {[\theta^4_3(\bar\tau)-\theta^4_4(\bar\tau)-\theta^4_2(\bar\tau)]
   \over \eta^4(\bar\tau)},
\ee
where the first factor refers to ten dimensional Minkowski space in light cone
gauge, the part holomorphic in the complex world sheet parameter $\tau$
describes the gauge part of the left handed bosonic string and the
antiholomorphic part account for the right handed superfermions.
We introduced Dedekind's $\eta$-function
\be \label{dedekind}
   \eta(\tau)=q^{1/24} \prod_{n=1}^{\infty} (1-q^n)
\ee
and Jacobi's $\theta$-functions
\be
   \th \left[ \ba{c} \alpha \\ \beta \ea \right] (\tau) =
       \sum_{n \in {\bf Z}} q^{\halb (n+\alpha)^2}e^{2\pi i(n+\alpha)\beta}
\ee
with $q=e^{2 \pi i \tau}$.
In the context of toroidal orbifolds each part can be realized as free
world sheet bosons. Thus the $\theta$-functions will be regarded as
the instanton part whereas the $\eta$-functions describe quantum
oscillations. Factors like $(2 \pi \imag \tau )$ arise after integrating out
continuous momentum states. After compactifying on $T^6$ this integration
transmutes to a sum over discrete momenta $p^\mu$ characterized by elements
of the dual torus lattice $\Lambda^\ast$. Modular invariance requires the
appearance of an additional sum over elements of $\Lambda$ itself
interpreted as winding states $w^{\mu}$. These two lattices can be combined
to define an even self-dual lattice $\Lambda_N$ with Lorentzian
signature~\cite{Narain}
$$ (+,+,+,+,+,+,-,-,-,-,-,-) $$
and elements $P=(P_L,P_R)$, defined in (\ref{prpl}).

Again we used the freedom of turning on an antisymmetric background field
$b_{ij}$ which corresponds to considering {\em all\/} even, self-dual lattices
in $6+6$ dimensions\footnote{Similarly, it is possible to add the
$E_8 \times E_8$ gauge lattice as well and to allow for the most general
lattice in $22 + 6$ dimensions of even and self-dual type, what in turn
corresponds to turning on Wilson lines.}. The partition function now reads
\be \label{ZTorus}
   Z(\tau,\bar{\tau})= {\sum\limits_{P \in \Lambda_N}
   q^{P_L {1\over 2g} P_L} \bar{q}^{P_R {1\over 2g} P_R} \over
   8 (2 \pi \imag \tau ) |\eta(\tau)|^{16}}
   {[ \theta^8_3(\tau)+\theta^8_4(\tau)+\theta^8_2(\tau)]^2 \over
   \eta^{16}(\tau)}
   {[\theta^4_3(\bar\tau)-\theta^4_4(\bar\tau)-\theta^4_2(\bar\tau)]
   \over \eta^4(\bar\tau)}.
\ee
The orbifold projection can be performed yielding the {\em untwisted
partition function\/}
\be
\sum_{n=0}^{N-1} Z_{(1,\theta^n)}(\tau,\bar{\tau})=
\quart \sum_{n=0}^{N-1} {\rm Tr} \, \theta^n q^{L_0} \bar{q}^{\bar{L}_0},
\ee
where the trace is to be taken over all states of the torus
theory\footnote{not just the ones subject to {\em physical state
conditions}} with conformal dimensions $L_0$ and $\bar{L}_0$.
The projection is such that oscillator states in the quantum part
are multiplied by phases $\alpha = e^{2 \pi i/N}$ and powers thereof,
whereas instantons are organized in {\em orbit sums\/}
with definite twist eigenvalues. In other words, orbit sums are linear
combinations of states of toroidal Hilbert space diagonalizing the twist.
We conclude that $Z_{(1,1)}$ is simply
given by $1/N$ times the torus function and thus clearly contains the
full instanton part. If the defining twist matrix $\theta$ fixes the origin
of the lattice {\em only}, as is the case for all our models, $Z_{(1,\theta)}$
will {\em not\/} contain any instantons from the six dimensional
part, since the phases of the orbits add up to zero. Clearly, the same
is true for all $Z_{(1,\theta^n)}$, where $n$ is not a divisor of $N$.
In all other cases the appearance of an instanton sum precisely depends
on the question whether the corresponding power of $\theta$ leaves fixed
tori or not. Namely, if it does give rise to fixed directions the corresponding
instantons are (like the origin) not organized in orbit sums. It is this fact
where our disagreement with the literature stems from. For instance, in
formulae (3.3a) and (3.3c) of reference~\cite{IMNQ} there is no such sum.
In these formulae there appear correctly the instanton sums coming
from the gauge part. The remark to make is simply that instanton sums
can appear in parts of $Z_{(1,\theta^n)}$ even if the lattice is not invariant
under $\theta$ itself.

The twisted sectors of the orbifold can be obtained by successive
$S$ and $T$ trans\-formations\footnote{S: $\tau \rightarrow - 1/\tau$;
T: $\tau \rightarrow \tau + 1$.}, which then ensure one loop modular
invariance by construction~\cite{Ginsparg}. For symmetric
$Z_N$-orbifolds these
transformations close after having created $N(N-1)$ new terms labeled by
$Z_{(\theta^m,\theta^n)}$, with $1 \leq m < N$ and $0 \leq n < N$.
For these world sheet modular transformations we will need the identities
\bea
   \th \left[ \ba{c} \alpha \\ \beta \ea \right] (-1/\tau) &=&
   e^{2\pi i\alpha\beta} \sqrt{-i\tau} \hq
   \th \left[ \ba{c} \beta \\ -\alpha \ea \right] (\tau), \vier
   \th \left[ \ba{c} \alpha \\ \beta \ea \right] (\tau + 1) &=&
   e^{i\pi\alpha (1 - \alpha)} \hq
   \th \left[ \ba{c} \alpha \\ \alpha + \beta - \halb\ea \right] (\tau).
\eea

The question arises whether the invariant lattice
we had argued for in the last paragraph has implications on the spectrum
of twisted states and in particular on the {\em generalized GSO-projection\/}
established in~\cite{LMN}. Concerning massive states the answer is
certainly yes, since the lattice dual to the invariant lattice contributes.
Massless states are never built up by states of this dual lattice, so
how can the number of massless matter multiplets depend on it? The important
point is that the invariant lattice lowers the degeneracy factors of
$Z_{(\theta^m,\theta^n)}$ iff its volume differs from one. As a corollary of
our classification in the proceeding section we can state that for
any configuration of twist eigenvalues there exist a model where
the volume of the invariant lattice is one. In the remaining 9 cases
$h_{1,1}$ is reduced.

We don't need to worry whether we really obtain a sensible string theory
after dividing the degeneracy factors by the volume
of the invariant sublattice, i.e.\ whether we get an integer number of states.
As shown in~\cite{Lepo,NSV} this is guaranteed due to the fact that
$\Lambda_N$ appearing in (\ref{ZTorus}) is even and self-dual.

In order to illustrate all this, we now discuss as an example the $Z_4$ case
in more detail. In particular, we compare models 2 and 4 in our list. These
are also the most explicit ones of reference~\cite{CGM}, where for the
first time such a comparative study of different lattices in a somewhat
different context was undertaken.

First we give the partition functions.
In order to keep the formulae readable we will now restrict ourselves to the
twisted bosons of internal
space. I.e., the holomorphic gauge and antiholomorphic superfermionic parts
as well as uncompactified space time dimensions are disregarded and can
be found in~\cite{IMNQ}.
$$ \label{untwist} \ba{ll}
   Z_{(1,1)}^{int.} = \ds{{\sum\limits_{P \in \Lambda_N}
               q^{P_L^2} \bar{q}^{P_R^2} \over
               \big| \eta(\tau) \big|^{12}}} \;\; &
   Z_{(1,\th)}^{int.}=16\ds{{ \big|\eta(\tau)\big|^6 \over \Bigl|
        \th^2 \Bigl[ \ba{c} \sss{1/2} \\ \sss{3/4}\ea \Bigr] \, \,
        \th \Bigl[ \ba{c} \sss{1/2} \\ \sss{0}  \ea \Bigr] \Bigr|^2}} \\
   Z_{(1,\th^2)}^{int.}=16\ds{{ \sum\limits_{P \in \Lambda_N^\bot}
      q^{P_L^2} \bar{q}^{P_R^2} \over \Bigl|
      \th^2 \Bigl[ \ba{c} \sss{1/2} \\ \sss{0} \ea \Bigr] \Bigr|^2}} \;\; &
   Z_{(1,\th^3)}^{int.}=16\ds{{ \, \big|\eta(\tau)\big|^6 \over
        \Bigl| \th^2 \Bigl[ \ba{c} \sss{1/2} \\ \sss{1/4}\ea \Bigr] \, \,
       \th \Bigl[ \ba{c} \sss{1/2} \\ \sss{0}  \ea \Bigr] \Bigr|^2}} \\[ 6 mm]

   Z_{(\th,1)}^{int.}=16\ds{{ \, \big|\eta(\tau)\big|^6 \over \Bigl|
        \th^2 \Bigl[ \ba{c} \sss{3/4} \\ \sss{1/2}\ea \Bigr] \, \,
        \th \Bigl[ \ba{c} \sss{0} \\ \sss{1/2}  \ea \Bigr] \Bigr|^2}} \;\; &
   Z_{(\th,\th)}^{int.}=16\ds{{ \big|\eta(\tau)\big|^6 \over \Bigl|
        \th^2 \Bigl[ \ba{c} \sss{3/4} \\ \sss{3/4}\ea \Bigr] \, \,
        \th \Bigl[ \ba{c} \sss{0} \\ \sss{0}  \ea \Bigr] \Bigr|^2}} \\[ 6 mm]
   Z_{(\th,\th^2)}^{int.}=16\ds{{ \big|\eta(\tau)\big|^6 \over
        \Bigl| \th^2 \Bigl[ \ba{c} \sss{3/4} \\ \sss{0}\ea \Bigr] \, \,
        \th \Bigl[ \ba{c} \sss{0} \\ \sss{1/2}  \ea \Bigr] \Bigr|^2}} \;\; &
   Z_{(\th,\th^3)}^{int.}=16\ds{{ \big|\eta(\tau)\big|^6 \over \Bigl|
        \th^2 \Bigl[ \ba{c} \sss{3/4} \\ \sss{1/4}\ea \Bigr] \, \,
        \th \Bigl[ \ba{c} \sss{0} \\ \sss{0}  \ea \Bigr] \Bigr|^2}} \\[ 6 mm]

   Z_{(\th^2,1)}^{int.}=16\ds{\sum\limits_{P \in (\Lambda_N^\bot)^\ast}
   q^{P_L^2} \bar{q}^{P_R^2}\over{{\rm vol\;} \Lambda_N^\bot}\;\;
   \Bigl| \th^2 \Bigl[ \ba{c} \sss{0} \\ \sss{1/2} \ea \Bigr] \Bigr|^2} \;\; &
   Z_{(\th^2,\th)}^{int.}=16\ds{\big|\eta(\tau)\big|^6 \over
        \Bigl| \th^2 \Bigl[ \ba{c} \sss{0} \\ \sss{1/4}\ea \Bigr] \, \,
        \th \Bigl[ \ba{c} \sss{1/2} \\ \sss{0}  \ea \Bigr] \Bigr|^2} \\[ 6 mm]
   Z_{(\th^2,\th^2)}^{int.}=16\ds{{\sum\limits_{P\in(\Lambda_N^\bot)^\ast}
   e^{2\pi i P_L^2 - P_R^2} \; q^{P_L^2} \bar{q}^{P_R^2}
   \over {{\rm vol}\; \Lambda_N^\bot}\;\;
   \Bigl| \th^2 \Bigl[ \ba{c} \sss{0} \\ \sss{1/2} \ea \Bigr] \Bigr|^2}} \;\; &
   Z_{(\th^2,\th^3)}^{int.}=16\ds{{\big|\eta(\tau)\big|^6 \over
        \Bigl| \th^2 \Bigl[ \ba{c} \sss{0} \\ \sss{3/4}\ea \Bigr] \, \,
        \th \Bigl[ \ba{c} \sss{1/2} \\ \sss{0}  \ea \Bigr] \Bigr|^2}} \\[ 6 mm]

   Z_{(\th^3,1)}^{int.}=16\ds{{\big|\eta(\tau)\big|^6 \over \Bigl|
        \th^2 \Bigl[ \ba{c} \sss{1/4} \\ \sss{1/2}\ea \Bigr] \, \,
        \th \Bigl[ \ba{c} \sss{0} \\ \sss{1/2}  \ea \Bigr] \Bigr|^2}} \;\; &
   Z_{(\th^3,\th)}^{int.}=16\ds{{\big|\eta(\tau)\big|^6\over \Bigl|
        \th^2 \Bigl[ \ba{c} \sss{1/4} \\ \sss{1/4}\ea \Bigr] \, \,
        \th \Bigl[ \ba{c} \sss{0} \\ \sss{0}  \ea \Bigr] \Bigr|^2}} \\[ 6 mm]
   Z_{(\th^3,\th^2)}^{int.}=16\ds{{\big|\eta(\tau)\big|^6 \over
        \Bigl| \th^2 \Bigl[ \ba{c} \sss{1/4} \\ \sss{0}\ea \Bigr] \, \,
        \th \Bigl[ \ba{c} \sss{0} \\ \sss{1/2}  \ea \Bigr] \Bigr|^2}} \;\; &
   Z_{(\th^3,\th^3)}^{int.}=16\ds{{\big|\eta(\tau)\big|^6\over\Bigl|
        \th^2 \Bigl[ \ba{c} \sss{1/4} \\ \sss{3/4}\ea \Bigr] \, \,
        \th \Bigl[ \ba{c} \sss{0} \\ \sss{0}  \ea \Bigr] \Bigr|^2}}
\ea $$
We denoted the invariant part of the self-dual lattice by $\Lambda_N^\bot$.
The numerical factors in the untwisted sector arise when expressing
infinite products in terms of $\theta$-functions. They carry over to the
twisted parts, but they are implicitly lowered whenever a function
$\th \Bigl[ \ba{c} \sss{1/2} \\ \sss{\beta}  \ea \Bigr]$
appears. In our case the actual degeneracy factor in $Z_{(\th^2,\th)}$
and $Z_{(\th^2,\th^3)}$ are thus reduced to 4. Recalling that the whole
partition function is multiplied by the inverse twist order this yields
projectors for {\em massless twisted states\/} of the form
(compare~\cite{IMNQ})
\bea
\quart (16 + 16 \Delta_\th + 16 \Delta_\th^2 + 16 \Delta_\th^3) \\
\quart ({16\over {{\rm vol}\; \Lambda_N^\bot}} + 4 \Delta_{\th^2} +
  {16\over {{\rm vol}\; \Lambda_N^\bot}}\Delta_{\th^2}^2 +
  4 \Delta_{\th^2}^3) \label{www} \\
\quart (16 + 16 \Delta_{\th^3} + 16 \Delta_{\th^3}^2 + 16 \Delta_{\th^3}^3),
\eea
where $\Delta=\pm 1$ for generations and antigenerations, respectively.
As mentioned earlier the only difference to the corresponding projector
in~\cite{IMNQ} is the appearance of the volume factors. It differs from one
exactly in the cases where $\Lambda_N^\bot$ does not factorize trivially.
For model 2 it is just given by a two-dimensional torus lattice and its dual.
Being self-dual such a lattice transforms into itself under the Poisson
resummation\footnote{The Poisson resummation formula reads
$$\sum\limits_{w \in \Lambda} \exp[-\pi (w+\epsilon)^T A (w+\epsilon) +
 2\pi i \phi^T (w+\epsilon)] = {1\over {\rm vol} \; \Lambda \;\sqrt{\det A}}
 \sum\limits_{p \in \Lambda^\ast} \exp[-\pi (p+\phi)^T A^{-1} (p+\phi) -
 2\pi i \epsilon^T p],$$ where $\epsilon$ and $\phi$ are constant vectors
 and $A$ is an invertible matrix.}

connected with the $S$ transformation and we conclude
${\rm vol}\; \Lambda_N^\bot = 1$.
Expressed in more formal terms for the torus lattice we have the relation
\be \label{xxx}
   (\Lambda^\ast)^\bot = (\Lambda^\bot)^\ast
\ee
The same is true for the \underline{first} model of each $Z_N$ $(Z_N^\prime)$
in table (\ref{orbtab}), respectively. The corresponding values $h_{(1,1)}$ and
$h_{(1,2)}$ are the ones stated in the literature. In contrast, the other cases
no longer satisfy (\ref{xxx}) as will now be illustrated with help of model 4.

The $Z_4$-twist matrix and its dual in an $SU(4)$-lattice are
\be
\th = \left( \ba{rrr} 0 & 0 & -1 \\ 1 & 0 & -1 \\ 0 & 1 & -1 \ea \right),\;\;\;
\th^\ast = \left( \ba{rrr} -1 & -1 & -1 \\ 1 & 0 & 0 \\ 0 & 1 & 0 \ea \right) .
\ee
In solving the equations
\be
\th^2 n = n , \;\;\; {{\th^T}^{-1}}^2 m = m,
\ee
we find the fixed directions
\be \label{fix4}
(n_1,0,n_1), \;\;\; (m_1,-m_1,m_1).
\ee
To find the volume factor of $\Lambda_N^\bot$ in this case,
we consider one of the two scalar products in (\ref{ener}) or (\ref{spin}).
It is convenient to take the latter one, since it does not depend on the
background parameters.
We just compute the quantity $2 m^T n$ for the sublattices defined in
(\ref{fix4}) and normalize the result w.r.t.\ to a (self-dual) circle
theory,
\be
   {\rm vol} \; \Lambda_N^\bot = \frac{4 m_1^T n_1}{2 m_1^T n_1}.
\ee
Since model 4 is a product of two $SU(4)$-lattices, we finally find
a volume factor of four and we can read off from the projector in (\ref{www}),
that four generations and no antigeneration survive in the double twisted
sector of this model. Comparing with model 2, where ten generations and
six antigene\-ra\-tions survive the projection, we make two observations which
in fact turn out to be general rules. First, the number of {\em chiral\/}
generations is unchanged when considering different lattices but the same
twist eigenvalues. This one indeed expects, since this number can be
computed by a formula conjectured by Dixon, Harvey, Vafa and Witten~\cite{DHVW}
and proved by Markushevich, Olshanetsky and Perelomov~\cite{MOP}, which only
uses twist eigenvalues. Second, the total number of generations is lowered
when one considers cases where formula (\ref{xxx}) is no longer fulfilled.
For the projector of massive states
the phase factor $e^{2\pi i P_L^2 - P_R^2}$ in $Z_{(\th^2,\th^2)}^{int.}$
has to be taken into account in case of non-trivial instanton contributions.

The list of the basic degeneracy factors for all twist eigenvalue
confi\-gu\-rations can be found in~\cite{KKKOOT}. Here, we
give the volume factors $V_i$ for all our models, where the subscript
denotes the sector in which the fixed torus appears.

\vskip  .3 cm
\centerline{Table 4: Volume factors $V_i$ for twist sectors with fixed tori.}
\vskip  .3 cm
\begin{center}
\begin {tabular}{c|c}
   model & volume factors \\ \hline
     1   &       ---       \\
     2   &       $V_2=1$   \\
     3   &       $V_2=2$   \\
     4   &       $V_2=4$   \\
     5   &       $V_3=1$   \\
     6   &       $V_3=4$   \\
     7   &   $V_2=1 \;\;\;\;\; V_3=1$   \\
     8   &   $V_2=1 \;\;\;\;\; V_3=4$   \\
     9   &   $V_2=3 \;\;\;\;\; V_3=1$   \\
     10  &   $V_2=3 \;\;\;\;\; V_3=4$   \\
     11  &       ---        \\
     12  &       $V_4=1$    \\
     13  &       $V_4=4$    \\
     14  &   $V_2=1 \;\;\;\;\; V_4=1$    \\
     15  &   $V_2=2 \;\;\;\;\; V_4=2$    \\
     16  &   $V_3=1 \;\;\;\;\; V_6=1$    \\
     17  &   $V_3=4 \;\;\;\;\; V_6=4$    \\
     18  &   $V_2=1 \;\;\;\;\; V_4=1 \;\;\;\;\; V_6=1$
\end {tabular}
\end{center}

For the resulting generation numbers we refer to table~\ref{orbtab}.

\section{Geometrical Resolution of the Orbifold Singularities}
\label{geometry}

In this section we will calculate the Hodge numbers for the Calabi-Yau
manifold which is constructed by resolving the orbifold singularities.
As it was conjectured in \cite{DHVW}
there exists a resolution $\widehat {T/G}$ of the toroidal
orbifold $T/G$ to a Calabi-Yau manifold,
if the group action leave the holomorphic three form of the torus invariant.
The prediction for the Euler number was extracted from the partition
function \cite{DHVW} and stated in form of the famous orbifold formula
\be \chi(\widehat {T/G}) = {1\over |G|} \sum_{[g,h]=0} \chi_{g,h},
\label{eulernumber}
\ee
where the sum is taken over all commuting elements of the group and
$\chi_{g,h}=\chi(Fix(g) \cap Fix(h))$ is the Euler number of
the intersection of the fixed sets under $g$ and $h$, respectively.
For the special case of $Z_N$ actions the corresponding
Calabi-Yau spaces were constructed explicitly \cite{MOP,RY} confirming
(\ref{eulernumber}).

\subsection{The fixed sets}
\label{fixsets}
In our examples we have only fixed points $P$ ($\chi(P)=1$) and fixed
tori $T$ ($\chi(T)=0$).  The application of
(\ref{eulernumber}) is simplified by the fact that
$\chi_{g,h}=\chi(Fix(gh))$. Because $\chi(T)=0$ the
precise number of fixed tori in the higher sectors does not
affect the Euler number. Using the numbers of fixed points which
depend only on the eigenvalues of the twist matrix (cf. section
\ref{class} table 2) we see that $\chi=72$ in the $Z_3$ case and
$\chi=48$ in all other cases.

In contrast to the Euler number the numbers of independent (1,1)-forms
and (1,2)-forms in Hodge cohomology depend on the fixed tori
and their intersection pattern and hence on the lattice.
Let us illustrate this point with help of a series of examples, namely
three $Z_6'$ models with twist exponents ${1\over 6}(1,2,3)$.
As the first model we consider case 7 of table \ref{orbtab}
which is equivalent to the Coxeter twist in the lattice
$A_1\times A_1\times A_2\times G_2$.  An explicit twist matrix
can be obtained by combining the irreducible blocks
$(Z_2^{(1)},Z_2^{(1)},Z_6^{(2)},Z_3^{(2)})$
specified by the vectors in table 3.
This matrix has the following fixed sets in the first, second and
third twist sector respectively:

\begin{tabbing}
$\theta$:\phantom{xxx}\= 12 fixed points: \= $ v_i\otimes (0,0)\otimes w_j$,
$i=1,2,3,4$;  $j=1,2,3$ \\
\> \>
$v_1=(0,0),\,\,
v_2=(\frac{1}{2},0),\,\,
v_3=(0,\frac{1}{2}),\,\,
v_4=(\frac{1}{2},\frac{1}{2})$;\\
\> \>
$w_1=(0,0),\,\,
w_2=(\frac{1}{3},\frac{2}{3}),\,\,
w_3=(\frac{2}{3},\frac{1}{3})$ \\
$\theta^2$: \> 9 fixed tori: \> invariant subspace: $n_1=(1,0,0,0,0,0),\,
n_2=(0,1,0,0,0,0)$; \\
\> \> \phantom{invariant subspace:} $[m_1=(1,0,0,0,0,0),\, m_2=(0,1,0,0,0,0)$]
\\
\> \> basepoints: \= $(0,0)\otimes v_i\otimes w_j$, $i,j=1,2,3$\\
\> \> \>
$v_1=(0,0),\,\,
v_2=(\frac{1}{3},\frac{1}{3}),\,\,
v_3=(\frac{2}{3},\frac{2}{3})$;\\
\> \> \>
$w_1=(0,0),\,\,
w_2=(\frac{1}{3},\frac{2}{3}),\,\,
w_3=(\frac{2}{3},\frac{1}{3})$\\
$\theta^3$: \> 16 fixed tori: \> invariant subspace: $n_1=(0,0,1,0,0,0),
n_2=(0,0,0,1,0,0)$;\\
\> \> \phantom{invariant subspace:} $[m_1=(0,0,1,0,0,0),\,
m_2=(0,0,0,1,0,0)$]\\
\> \> base points: \= $v_i\otimes (0,0)\otimes v_j$, $i,j=1,2,3,4$\\
\> \> \>
$v_1=(0,0),\,\,
v_2=(\frac{1}{2},0),\,\,
v_3=(0,\frac{1}{2}),\,\,
v_4=(\frac{1}{2},\frac{1}{2})$\\
\end{tabbing}
The vectors $n_i$ span an invariant subspace of the matrix $\theta^s$
from the given base points. These data define the corresponding fixed tori.
The vectors $m_i$ span the invariant subspace of
$\left({{\theta^T}^{-1}}\right)^s$; their
significance is explained below.
Note that all base points given, lie in fact
on different tori, as their only non vanishing entries
are perpendicular to the $n_i$.
The schematic view of the intersection pattern can be found in
figure 1. Let us compare this situation
with the one of case 8, which may be defined as the the Coxeter twist in the
lattice $A_2\times D_4$
\be \theta=\left(
\begin{array}{rrrrrr}
\phantom{-}0&
-1&
\phantom{-}0&
\phantom{-}0&
\phantom{-}0&
\phantom{-}0 \\
\phantom{-}1&
-1&
\phantom{-}0&
\phantom{-}0&
\phantom{-}0&
\phantom{-}0 \\

\phantom{-}0&
\phantom{-}0&
\phantom{-}0&
\phantom{-}1&
-1&
-1 \\

\phantom{-}0&
\phantom{-}0&
\phantom{-}1&
\phantom{-}1&
-1&
-1 \\

\phantom{-}0&
\phantom{-}0&
\phantom{-}0&
\phantom{-}1&
-1&
\phantom{-}0 \\

\phantom{-}0&
\phantom{-}0&
\phantom{-}0&
\phantom{-}1&
\phantom{-}0&
-1
\end{array} \right)  \label{case 8}\ee
and has the following fixed sets
\begin{tabbing}
$\theta$:\phantom{xxx}\= 12 fixed points: \= $ v_i\otimes w_j$,
$i=1,2,3$;  $j=1,2,3,4$ \\
\> \>
$v_1=(0,0),\,\,
v_2=(\frac{1}{3},\frac{2}{3}),\,\,
v_3=(\frac{2}{3},\frac{1}{3})$\\
\>\>
$w_1=(0,0,0,0),\,\,
w_2=(0,0,\frac{1}{2},\frac{2}{2}),\,\,
w_3=(\frac{1}{2},0,0,\frac{1}{2}),\,\,
w_4=(\frac{1}{2},0,\frac{1}{2},0)
$\\
$\theta^2$: \> 3 fixed tori: \> invariant subspace: $n_1=(0,0,1,0,0,1),\,
n_2=(0,0,1,0,1,0)$;\\
\> \> \phantom{invariant subspace:} $ [m_1=(0,0,1,-1,0,1),\,
m_2=(0,0,1,-1,1,0)$]\\
\> \> basepoints: \= $v_i\otimes (0,0,0,0)$, $i=1,2,3$\\
\> \> \>
$v_1=(0,0),\,\,
v_2=(\frac{1}{3},\frac{2}{3}),\,\,
v_3=(\frac{2}{3},\frac{1}{3})$\\
$\theta^3$: \> 16 fixed tori: \> invariant subspace: $n_1=(1,0,0,0,0,0),\,
n_2=(0,1,0,0,0,0)$;\\
\> \> \phantom{invariant subspace:} $[m_1=(1,0,0,0,0,0),\, (0,1,0,0,0,0)$]\\
\> \> basepoints: \= $(0,0)\otimes v_i\otimes v_j$, $i,j=1,2,3,4$\\
\> \> \>
$v_1=(0,0),\,\,
v_2=(\frac{1}{2},0),\,\,
v_3=(0,\frac{1}{2}),\,\,
v_4=(\frac{1}{2},\frac{1}{2})$\\
\end{tabbing}
This model was investigated in detail in \cite{A}, however here
the author claims that there are $9$ instead of  $3$ fixed tori
in the second twisted
sector and the conclusion about the massless spectrum
is  therefore not completly correct.
The reason for the difference seems to be due to an improper use
of the {\sl Lefschetz Fixed Point Theorem}.
In the second twist sector the exponents of $\theta_d^2$ are
${1\over 3}(1,2,0)$, so that the coordinate plane $\tilde I$
spanned by $z_3$ (\ref{co-ord}), is fixed. In order to
calculate the multiplicity of the corresponding fixed tori,
the authors of \cite{MOP,A} apply now the {\sl Lefschetz Fixed Point Theorem}
(\ref{lft}) to the action of $\theta_d^2$ on the subspace $\tilde J$,
spanned by the first and second coordinate plane.
The result is $n_F={\rm det}(1-\theta^2)|_{\tilde J}=$ $(1-\exp[2\pi i/3])^2
(1-\exp[4\pi i/3])^2=9$.
This is inadquate, because the splitting of $\IR^6$
in $\tilde I$ and $\tilde J$ does not correspond to a splitting of the
lattice $\Lambda$ into sublattices on which $\theta^2$ acts as an
automorphism.
Let us  pass to the lattice basis and denote the sublattice fixed
w.r.t.\ to the lattice automorphism  $\theta^2$ by $I$.
The {\sl Lefschetz Fixed Point Theorem} could be utilized in the
above sense, if there would be a sublattice $J$ invariant under $\theta^2$,
which is complementary to $I$, i.e.\ $\Lambda=I\oplus_\ZZ J $.
This is not the case because $\theta^2$ has no block structure w.r.t.\ $I$.

In the third twisted sector $\theta^3$ has block structure w.r.t
its invariant sublattice $I$ so the reasoning of the authors
\cite{MOP,A} yields the correct result.
Similarly it applies to the second and third twisted sector of example 7.

Instead of calculating the fixed sets explicitly the
{\sl Fixed Point Theorem} can be modified by taking into account
volume factors in the Poisson resumation formula
which reduces the multiplicity of the twisted states as it was
explained in the previous section.
Let $\Theta$ be the action of the twist in the Narain lattice
labeled by $(n,m)$ and $I$ the invariant subspace in this lattice.
The number of connected fixed sets is then given by~\cite{NSV}
\be n_F=\sqrt{
{{\rm det'}({\bf 1}-\Theta)} \over {{\rm Vol}_{N\,}} [I] },\label{glft}
\ee
where the evalution of ${\rm det'}$ is defined by taking the product over
the nonzero eigenvalues only. The volume is with respect to the
Narain scalar product (\ref{spin}) and normalized such that the unit cell
has volume $1$. In order to cover also the case of fixed points
we define ${\rm Vol}_N$ of any number of discrete points to be $1$.
In  the cases at hand we have
$$\Theta=\pmatrix{\theta & 0\cr 0&{\theta^T}^{-1}}.$$
Let $m_i$ be the vectors which span the sublattice invariant under
${\theta^T}^{-1}$. The formula (\ref{glft}) simplifies to
\be n_F={{\rm det'}({\bf 1}-\theta) \over {\rm det}'(n_i^T m_j)};\label{sglft}
\ee
e.g. for the $\theta^2$ sector of the case above we
get indeed the reduction factor ${\rm det}'(n_i^T m_j)=3$. For the other
volume factors see table 4.

As the last example in our series we consider the case $10$
which can be realised as the Coxeter twist in $A_1\times A_5$. Our
canonical twist $(Z_2^{(1)}, Z_6^{(5)})$ has the following fixed
sets (cf. figure 1)

\begin{tabbing}
$\theta$:\phantom{xxx}\= 12 fixed points: \= $ v_i\otimes w_j$,
$i=1,2$;  $j=1,2,3,4,5,6$ \\
\> \>
$v_1=(0),\,\,
v_2=(\frac{1}{2})$\\
\> \>
$w_1=(0,0,0,0,0)\,\,
w_2=\frac{1}{6}(1,2,3,4,5),\,\,
w_3=\frac{1}{6}(2,4,0,2,4),$\\
\> \>
$w_4=\frac{1}{6}(3,0,3,0,3),\,\,
w_5=\frac{1}{6}(4,2,0,4,2),\,\,
w_6=\frac{1}{6}(5,4,3,2,1)
$\\
$\theta^2$: \> 3 fixed tori: \> invariant subspace: $n_1=(0,1,0,1,0,1),\,
n_2=(1,0,0,0,0,0)$;\\
\> \> \phantom{ invariant subspace: } $[m_1=(0,1,-1,1,-1,1),\,
m_2=(1,0,0,0,0,0)$]\\
\> \> base points: \= $(0,0) \times v_i$, $i=1,2,3$\\
\> \> \>
$v_1=(0,0,0,0),\,\,
v_2=(\frac{1}{3},\frac{1}{3},\frac{2}{3},\frac{2}{3} ),\,\,
v_3=(\frac{1}{3},\frac{1}{3},\frac{2}{3},\frac{2}{3})$\\
$\theta^3$: \> 4 fixed tori: \> invariant subspace: $n_1=(0,1,0,0,1,0),\,
n_2=(0,0,1,0,0,1)$;\\
\> \> \phantom{ invariant subspace: } $[m_1=(0,1,0,-1,1,0),\,
m_2=(0,0,1,-1,0,1)$]\\
\> \> base points: \= $ v_i\otimes w_j$, $i,j=1,2$\\
\> \> \>
$v_1=(0),\,\,
v_2=(\frac{1}{2})$,\\
\> \> \>
$w_1=(0,0,0,0,0),\,\,
w_2=(\frac{1}{4},\frac{1}{4},\frac{1}{2},\frac{3}{4},\frac{3}{4})$\\
\end{tabbing}
The fixed sets where e.g. calculated in the appendix of
\cite{CGM}, however here  the conclusion was again that there
are $9$ tori in the second twisted sector and $16$ fixed
tori in the third twisted sector. The authors
have correctly calculated the vectors $n_i$ of the invariant lattice,
but as one can check, from the $9$ $(16)$ base points they give,
groups of $3$  $(4)$  lie on the same torus, respectively.

\subsection{Description of the desingularisations}
Let us now count the numbers of (1,1)-form introduced
by the resolutions of the fixed points and the fixed tori singularities.
By {\em Poincar\' e duality} it is equivalent to count the
{\em irreducible\/} components of the {\em exceptional divisors\/}, which are
introduced by the resolution process\footnote{See e.g.\ \cite{GH} as a
general reference for these concepts of algebraic geometry.}.
Below we review the necessary facts about the resolutions for the kind of
singularities we encounter.

In the case of fixed tori we have singularities due to the action of a
discrete $Z_k$ subgroup of $G$ on the $\IC^2$ fibres of the bundle normal
to the fixed tori $T^2$.
This action can locally always be recasted in the form
\be ( z_1, z_2 )\mapsto (\exp{2 \pi i q \over k} z_1,
\exp{2 \pi i \over k} z_2)
\ee
The singularity of $\IC^2/Z_k$ at the origin can be described by the
constraint
\be
y^k=x z^{1-q},
\ee
in $\IC^3$, where in our cases we have always $q=k-1$.
The resolution of this type of singularity, known as
the {\em rational double point} of type $A_{k-1}$, is
given~\cite{H} by a Hirzebruch-Jung
sphere tree. The number of spheres and their self intersection
numbers $-b_i$ in the resolved manifold can be obtained by an Euclidean
algorithm\footnote{The $b_i$
correspond to a representation of $k/q$ as continued fraction
of the following form
$${k\over q}=b_1-{1\over {b_2 -\ldots  {\displaystyle{-{1\over b_s}}} } }
.$$}~\cite{H}.
In the cases at hand\footnote{
The finite subgroups $G$ of $SU(2)$ fall into an
$ADE$ classification.
For this groups the intersection patterns of the spheres in the resolutions
of the singularities $\IC^2/G$ correspond to the
Dynkin diagrams, where points represent $\IP's$ and links represent
intersections,
i.e.\ the intersection matrix equals the negative
Cartan matrix.}
one gets a sequence of $k-1$ projective spaces
$\IP_1\vee \ldots \vee \IP_{k-1}$ joined
in one point with self intersection numbers
$\IP_i\cap\IP_{i}=-2$.
The resolution replaces the fibers in the normal bundle
over a {\em generic point} on the fixed
torus with a sphere tree. It introduces an {\em exceptional divisor}
of the form $T\times (\IP_1\vee \ldots \vee \IP_{k-1})$. The {\em new}
$h_{1,1}$ forms correspond to the number of irreducible components
of these exceptional divisors, which is $k-1$.

In the case of fixed points the singularities are locally of the
form $\IC^3/G$. If $G$ is abelian  as in our examples {\em toric geometry\/}
is a suitable framework to describe $\IC^r/G$ singularities and
their resolutions. In this sense it allows for a generalisation of the
above treatment of the $A_{k-1}$ singularities to higher dimensions.
In order to avoid lengthy repetitions
we refer to the book of Oda \cite{O} and the appendix of
Markushevich in \cite{MOP}
for the precise definitions and proofs of the properties of
{\em convex rational  polyhedral cones}, {\em fans\/} and
{\em toric varieties}.

Let $N\simeq \ZZ^r$ be a
lattice in an $\IR$ vector space $V$,
which is the completion of
$N$ over $\IR$, i.e.\ $V=N\otimes_{\ZZ}\IR$,
and $n_1,\ldots,n_s$ elements of $N$. A {\em strongly convex rational
polyhedral cone} with {\em apex} at the origin $O$ is defined as
\be
\tau:=\rp n_1 + \ldots + \rp n_s=\{a_1 n_1+\ldots +a_s n_s|a_i\in \rp\},
\label{scpc}
\ee
where $\tau\cap (-\tau)={O}$.
Let $n^{(1)},\ldots,n^{(N)}$ be a set of generators of the {\sl semi group}
${\cal S}_\tau=(\tau\cap N)$,
i.e. every lattice site in $(\tau\cap N)$ can be reached
by a linear combination of the $n^{(i)}$ with positive integer coefficients
and let
\be \ba{rl}
a_{11} n^{(1)} +\ldots +a_{1N} n^{(N)} &=0\\
                               \cdots & \cdots \\
a_{t1} n^{(1)} +\ldots +a_{tN} n^{(N)} &=0
\ea
\label{rel}
\ee
be a maximal set of linear relations among
the $n^{(1)},\ldots,n^{(N)}$ over $\ZZ$.
We can define from the above data an {\em affine toric variety}
$X^{(V,N,\tau)}$ as follows
\be
X^{(V,N,\tau)}:=\{(z_1,\ldots,z_N)\in \IC^N|
F_i(z_1,\ldots, z_N)=0,i=1,\ldots,t\},
\ee
where the $F_i (z_1,\ldots, z_N)=0$ are {\em monomial equations} of the
form
\be
z_1^{a_{i1}}\cdot\ldots\cdot z_N^{a_{iN}}=1.
\label{meq}
\ee
\newtheorem{example}{Example}
\begin{example}
$N\simeq\ZZ^r$ is a lattice in $\IR^r$ spanned by
$n_i,\, i=1,\ldots,r$.
$\tau=\rp n_1 + \ldots +\rp n_r =(\IR^+_0)^r$. Then we have {\rm no}
relation of the type {\rm (\ref{rel})}
for the generators of the semi group so  $X^{(\IR^r,N,\tau)}=\IC^r$.
\end{example}
\begin{example}
Let $N\simeq \ZZ^2$ be a lattice and $\tau=\rp (k n_1-n_2)+\rp n_2$.
We have $n^{(1)}=n_1$, $n^{(2)}=n_2$ and $n^{(3)}= (k n_1-n_2)$
(compare figure 2 a.)), so
that  $X^{(\IR^2,N,\tau)}=\{ (x,y,z)\in \IC^3|x^k=yz\}$ is the
rational double point of type $A_{k-1}$ encountered above.
\end{example}
More generally \cite{MOP,O}
\begin{example}
Let $G$ be a {\rm finite abelian  group} acting on $\IC^r$. Elements
$g\in G$ of order $k$ map
$(z_1,\ldots,z_r)\mapsto
(\exp[({2 \pi i\over k}) a_1]\, z_1,\ldots,
 \exp[({2 \pi i\over k}) a_r]\, z_r)$.
We have $\IC^r/G=X^{(\IR^n,N,\tau)}$,
with $\tau=\left(\rp \right)^n$ and
the lattice is defined as $\ds{N=\bigcap_{g\in G} N_g}$,
where the lattice $N_g$ is spanned by the minimal set of
vectors $n=(n_{(1)},\ldots,n_{(r)})$ with
$$n_{(1)} a_1+\ldots + n_{(r)} a_r\equiv 0\,\,mod\, k.$$
\end{example}
The constructive power of toric geometry lies in the fact that one
can glue together affine toric varieties in a natural way.
For this it is  convenient to pass first to the dual cone.
Denote by $W=V^{\vee}$ the dual of $V$ and by $M$
the dual lattice to $N$ w.r.t.\  the bilinear form
\be
\langle \, ,\, \rangle :M\times N \rightarrow \ZZ.
\ee
The dual cone $\sigma=\tau^{\vee} $ is defined by
\be
\sigma := \{x\in W|\langle x,y \rangle \ge 0,\,\, \forall y\in \tau\}.
\ee
Defining
\be
X_\sigma=X_{(W,M,\sigma)}:=X^{(V,N,\tau)}
\ee
has the advantage that $X_{\sigma\cap\sigma'}=
X_\sigma\cap X_{\sigma'}$. This property, which does not hold for
$X^\tau= X^{(V,N,\tau)}$, allows to visualize gluing of toric varieties by
gluing of cones.
{\em Faces} $\phi$ of a cone $\sigma$
are subsets which are defined via an element of the dual cone $n_0$
\be
\phi := \{ y\in \sigma|\langle y,n_0 \rangle = 0 \}.
\ee
Faces of strongly convex rational cones are strongly convex rational cones.
{\em Fans} $\Delta$
are made by sticking together cones $\sigma$ such that
\begin{itemize}
\begin{enumerate}
\item Every face of any $\sigma$ is in $\Delta$.
\item Every intersection $\sigma \cap \sigma'$ is a face of
       $\sigma$ and $\sigma'$.
\end{enumerate}
\end{itemize}
Let now $X_\sigma=X^{(V,N,\tau)}$ and $X_{\sigma'}=X^{(V,N,\tau')}$
be affine varieties associated to cones $\sigma,\sigma'\in \Delta $ and
$n^{(1)},\ldots,n^{(N)}$,  $n^{\prime (1)},\ldots,n^{\prime (N)}$,
generators of the semi groups
${\cal S}_\sigma$ and ${\cal S}_{\sigma'}$ respectively.
The transition functions between affine
coordinates $z_i$ and $z'_i$ are defined
by the maximal set of linear relations $j=1,\ldots ,s$
between these  generators
$$l_{1j}\, n^{(1)}+ \ldots +l_{Nj}\, n^{(N)}+l'_{1j}\, {n^\prime}^{(1)}+
\ldots +l'_{Nj}\,
{n^\prime}^{(N)}=0,$$
over $\ZZ$, as
\be z_1^{l_{1j}} \cdot \ldots \cdots z_n^{l_{Nj}} {z_1^\prime}^{l^\prime_{1j}}
\cdot \ldots \cdots {z_{N}^\prime}^{l^\prime_{Nj}}=1.\label{trf}\ee
Using this transition functions one can associate to a fan $\Delta$ a
{\em general toric variety}
$X_\Delta:=\ds{\bigcup_{\sigma\in \Delta} X_\sigma}$.

\begin{example}
Let $M\simeq\ZZ^2$ be lattice spanned by in $\IR^2$.
The fan $\Delta$ made by sticking together
$\sigma_1=\rp m_1+\rp m_2$,  $\sigma_2=\rp(-m_1 -m_2)+\rp m_2$ and
$\sigma_2=\rp(-m_1 -m_2)+\rp m_1$ defines $X_\Delta=\IP^2$.
The affine planes ($\{z_i\neq 0\}$ in homogeneous coordinates) correspond
to the $X_{\sigma_i}$. \end{example}
Note that the fan above covers the whole two plane.
As shown e.g. in Theorem 1.11. of \cite{O} we have in general
that $X_\Delta$ is compact if and only if $\Delta$ covers $W$.

Crucial for the resolution of singularities is the following
\newtheorem{toric}{Theorem}
\begin{toric}
The toric variety $X_\Delta$ associated  to a fan $\Delta$ in
$M\simeq \ZZ^r$ is nonsingular if and only if
each $\sigma\in \Delta$ is nonsingular in the following sense:
$\exists$ a $\ZZ$ basis $\{m_1,\ldots,m_r\}$ of $M$ such that
$\sigma=\rp m_1+\ldots + \rp m_s$. Such cones are called basic cones.
\label{tor1}
\end{toric}
A proof can be found in \cite{O}. If we consider the lattice
$M=\ZZ^2$ and the cone $\sigma=\rp m_1+\rp (k m_2+m_2) $
with $N=M^{\vee}=M$ and $\tau=\sigma^{\vee}$
as in Example 2, one sees that one has to {\em subdivide} $\sigma$
into a fan of $k$ {\em basic\/} cones in order to meet the requirement of
Theorem
\ref{tor1}.
The $k-1$ inner faces spanned by $\alpha_i$
by which this is achieved correspond to the $k-1$ exceptional $\IP$ curves
necessary to resolve the $A_{k-1}$  singularity, see figure 2 b.) below.

\vskip  4 mm
\centerline{Figure 2: Resolution of the $A_{k-1}$ rational double point
(k=3):}
\vskip  2 mm
\setlength{\unitlength}{ 12 mm}
\vskip  -1 mm
\begin{picture}(12,6)(-1,-1)
\put(1,1){\vector(1,0){1}}
\put(1.6,1.2){$n_1=n^{(1)}$}
\put(-.5,-,5){a.) The cone $\tau$}
\put(1,1){\vector(0,1){1}}
\put(-.5,1.6){$n_2=n^{(2)}$}
\put(1,1){\line(0,1){3}}
\put(1,1){\vector(3,-1){3}}
\put(2.4,-.2){$n^{(3)}$}
\put(1,1){\circle{.08}}
\put(1,2){\circle{.08}}
\put(1,3){\circle{.08}}
\put(1,4){\circle{.08}}
\put(4,0){\circle{.08}}
\put(2,1){\circle{.08}}
\put(2,2){\circle{.08}}
\put(2,3){\circle{.08}}
\put(2,4){\circle{.08}}
\put(3,1){\circle{.08}}
\put(3,2){\circle{.08}}
\put(3,3){\circle{.08}}
\put(3,4){\circle{.08}}
\put(4,1){\circle{.08}}
\put(4,2){\circle{.08}}
\put(4,3){\circle{.08}}
\put(4,4){\circle{.08}}

\put(7,-,5){b.) Subdivision of $\sigma$}
\put(8,1){\vector(1,0){1}}
\put(8.6,1.2){$m_1=m^{(1)}$}
\put(8,1){\line(1,0){1}}

\put(8,1){\vector(0,1){1}}
\put(8,1){\line(1,0){3}}
\put(8.1,3.5){$m^{(2)}$}
\put(7,1.6){$m_2$}
\put(8,1){\vector(1,3){1}}
\put(8,1){\vector(1,2){1}}
\put(9.2,2.8){$\alpha_2$}
\put(8,1){\vector(1,1){1}}
\put(9.2,1.8){$\alpha_1$}
\put(8,1){\line(1,3){1.2}}
\put(8,1){\line(1,2){1.3}}
\put(8,1){\line(1,1){1.5}}

\put(8,1){\circle{.08}}
\put(9,1){\circle{.08}}
\put(9,2){\circle{.08}}
\put(9,3){\circle{.08}}
\put(9,4){\circle{.08}}
\put(10,1){\circle{.08}}
\put(10,2){\circle{.08}}
\put(10,3){\circle{.08}}
\put(10,4){\circle{.08}}
\put(11,1){\circle{.08}}
\put(11,2){\circle{.08}}
\put(11,3){\circle{.08}}
\put(11,4){\circle{.08}}
\label{A2}
\end{picture}
\vskip  -2 mm

We are interested in resolution of singularities of type $\IC^3/Z_k$,
which have trivial canonical bundle. From this requirement we get
the following restriction \cite{MOP}.
\begin{toric}
Let $X_{W,M,\sigma}$ be a toric variety. Let $\sigma_i$ $i=1,\ldots,r$
a subdivision of $\sigma$ in the sense of Theorem 1. The
canonical bundle of $X_\Delta=
X_{\displaystyle{W,M,\cup_{i=1}^r \sigma_i}}$
is trivial if and only if the generators of ${\cal S}_{\sigma_i}$
lie on a hyperplane $H_N$ in $W$.
\label{tor2}
\end{toric}
Moreover the additional lattice vectors $\alpha_i$, which are
needed to define the subdivision, can be associated in a one to one
way to the compact divisors on $X_\Delta$.

Now consider the $Z_k$ actions whose exponents, see table 2, are of the form
${1\over k} (1,a_2,a_3)$ with
\be
1+a_2+a_3=0 \, {\rm mod}\, k.
\label{susy}\ee
In accordance with Example 3 we define the lattices $N,M$ by
\be
N=\langle n_1,n_2,n_3 \rangle_\ZZ =
                            \left\langle\ba{ccc}
                             k& -a_2& -a_3\\
                              0&    1&    0\\
                              0&    0&    1 \ea
\right\rangle_\ZZ \! \! ,
\quad
M=\langle m_1,m_2,m_3 \rangle_\ZZ =
                           \left\langle\ba{ccc}
                            {1\over k}&  0&  0\\
                            {a_2\over k}   &    1&    0\\
                            {a_3\over k}   &    0&    1 \ea
\right\rangle_\ZZ\! \! .
\ee
The cone $\sigma=\langle
e_1,e_2,e_3 \rangle_{\rp}$ is spanned by an orthonormal vectors as:
\vskip  4 mm
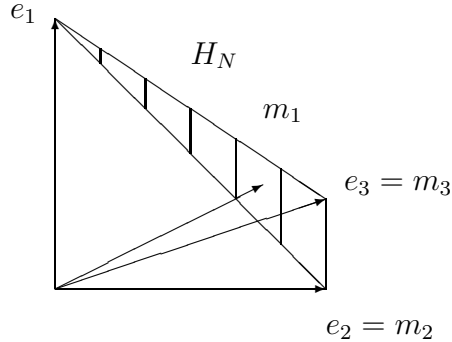
\begin{picture}(12,4)(1,0)

\put(6,1){\vector(0,1){3}}
\put(5.5,4){$e_1$}
\put(6,1){\vector(1,0){3}}
\put(9,0.5){$e_2=m_2$}
\put(6,1){\vector(3,1){3}}
\put(9.2,2.1){$e_3=m_3$}
\put(6,4){\line(1,-1){3}}
\put(6,4){\line(3,-2){3}}
\put(9,1){\line(0,1){1}}
\put(7.5,3.5){$H_N$}
\put(6.5,3.5){\line(0,1){.166}}
\put(7  ,3  ){\line(0,1){.333}}
\put(7.5,2.5){\line(0,1){.5}}
\put(8  ,2  ){\line(0,1){.666}}
\put(8.5,1.5){\line(0,1){.8333}}
\put(6,1){\vector(2,1){2.3}}
\put(8.3,2.9){$m_1$}
\label{A3}
\end{picture}
\vskip  2 mm
Due to (\ref{susy}) the lattice vector $m_1=\alpha_1$ lies indeed on the
hyper plane $H_N$. The other lattice vectors
$\alpha_i=(\alpha_i^{(1)},\alpha_i^{(2)},\alpha_i^{(3)})$, which define
the resolution according to Theorem 1 can be easily obtained as set of three
tuples :
\be \{k(\alpha_{i_j}^{(1)},\alpha_{i_j}^{(2)},\alpha_{i_j}^{(3)})=
(j\,  {\rm mod}\, k ,
j a_2\, {\rm mod}\, k ,
j a_3\, {\rm mod}\, k)| \sum_{l=1}^3 \alpha_{i_j}^{(l)}=k,j=1,\ldots,k-1\}.
\ee
We have drawn in figure 3 the trace of a resolving fan $X_\Delta$ in the
hyper plane $H_N$ for the $9$ different types of fixed point singularities.
To obtain a K\"ahlerian manifold we have also to make sure that
the corresponding resolutions are {\em projective algebraic}.
Indeed the  necessary conditions given in II.2.3 of \cite{O} are fulfilled.

The dots on the edges and in the interior of the the triangles $(e_1,e_2,e_3)$
correspond to $\alpha_i$ and hence to exceptional divisors of
$X_\Delta$. In the case of non prime
$k$  the fixed points of the $Z_k$ actions lie always on a fixed torus.
Note that the {\em exceptional divisors} associated to the points on the
edges of the triangle coincide  with the  exceptional divisors present over
the generic points of the fixed torus.
{}From figures 1 and 3 we can easily count the exceptional divisors.
E.g. consider the $Z_{12}'$ case 18.), here we have two tori fixed under
a group action of  order three, each gives rise to a
divisor $T\times(\IP\vee \IP)$, i.e. we get
$4$ (1,1)-forms from that resolution. The
resolutions of three fixed tori of order
two and one of order six add $3$ and $5$ (1,1)-forms, respectively.
The four fixed points of order four on the $Z_2$ fixed torus are
of type ${1\over 4} (1,1,2)$ and hence introduce $4$ {\em new }
exceptional divisors (cf. figure 3). Likewise the resolution of the
four $Z_{12}'$ fixed points on the $Z_6$ torus give rise to
$12$ exceptional divisors according to the inner points in the trace of the
last fan in figure 3.

A basis for the $9$ (1,1)-forms of the complex torus is given by
$dz_j\wedge d\bar z_j$ $i,j=1,2,3$ (\cite{GH} II.6). From this we see that
three (1,1)-forms are invariant if all twist exponents are different.
If two(three) twist exponents are equal we have five(nine)
invariant (1,1)-forms.
Adding in the above example the three invariant (1,1)-forms from the
torus we arrive $h_{1,1}=31$ the number given in table \ref{orbtab}.
Similarly the reader might use figures 2 and 3 to count all (1,1)-forms of all
other Coxeter orbifolds.
As $\chi= 2 (\chi_{1\, 1} - \chi_{2\, 1})$ for Calabi-Yau manifolds $h_{2\, 1}$
follows. Alternatively the additional $(1,2)$-forms can be obtained from
the properties of the resolving spaces \cite{O}.
Concerning the Hogde numbers we disagree in eight of the twelve cases given in
Table 1 of \cite {MOP}. We agree with the Hodge numbers given in
\cite{RY} for the prime cases $Z_3$ and $Z_7$.

While the  location of the the $\alpha_i$ is fixed by the
requirements of theorem 1 and 2, there
is some freedom in the triangulisation which lead to the actual choice
of the $\sigma_i$ and hence the resolving space.
Namely, whenever there is a quadruple of points
$\alpha_{i_1}\,\dots,\alpha_{i_4}$ in general position, the
diagonal of the quadrangle can be flipped by an {\em elementary
transformation}.

Given a triangulisation the triple intersection numbers of the irreducible
hypersurfaces can be calculated \cite{MOP}. Especially the intersection of
three exceptional divisors can be obtained by a simple algebraic
prescription \cite{roan}. Note that these numbers correspond to the
Yukawa couplings in the large radius limit. They can be used to check the
conformal field theory results for these couplings. The ambiguity in the
triangulisation process leads to different couplings. This ambiguity seems to
be hard in the sense that it cannot be removed by a field redefinition
\cite{A}.
{}From the point of view of conformal field theory it might correspond to an
ambiguity in taking the large radius limit in the orbifold moduli space.

\section{Conclusions}

We have investigated in a systematic way orbifolds
with $(2,2)$ world sheet supersymmetry, which can be constructed,
by modding out symmetric $Z_N$ actions from the six dimensional
torus with vanishing discrete $B$-field.
As a result we give a classification of these types of models.
Preceding investigations \cite{MOP,IMNQ,KKKOOT,CGM}
in this direction have the drawback of being
incomplete and more seriously of stating or using incorrect spectra.

The source for the deviations is that properties of the twist,
which are {\em not\/} directly related to the
twist eigenvalues, were not taken into account properly.
In the case of the $Z_4,Z_6,Z_6',Z_8,Z_8'$ and $Z_{12}$
actions there exist, for the same twist eigenvalues, inequivalent
automorphisms which are realised in different lattices.
The complete reducibility of the twist over $\IC$,
which is used to define the complex planes in the space-time basis,
is often confused with the reducibility of the twist over $\ZZ$, such that
the authors erroneously imply that the twist can be made block diagonal
in the lattice basis. Consequently their conclusions are only correct
if this holds indeed, which is usually {\em only\/} for one of the above
mentioned inequivalent automorphisms the case.
Statements concerning the factorisation properties of the modular
group, which have been made in the same spirit, are also wrong
in the general cases.

As the impact of the different lattices even on the spectrum
was not fully understood, almost all conclusions about these
non prime $Z_N$ orbifolds should be reconsidered. E.g. the couplings
stated in \cite{CGM} should be adapted to the correct spectrum etc.

Furthermore one should investigate how other
phenomenological relevant properties such as the
{\em non perturbative potential\/} for the moduli fields or the
{\em threshold corrections\/} to the gauge couplings will change,
when the different automorphisms with the same twist eigenvalues are
considered.

\section*{Acknowledgements}

\noindent
We are grateful to Hans-Peter Nilles and Dimitrios Dais for useful
comments and thank Luis Ib\'a\~nez for an email note.

\setcounter{table}{4}

\def\Z{\relax{\sf Z\kern-.4em Z}}
\small
\begin{table}
\caption{Hodge numbers of the $18$ symmetric $Z_N$-orbifolds
with $(2,2)$ world sheet supersymmetry and vanishing discrete
$b_{\mu\nu}$-field.
The corresponding twist eigenvalues are given in Table 2. The twist matrix
is specified by the irreducible blocks which appear in Table 3. We also
specify the Lie-algebra lattice where an equivalent twist can be realised
as generalised Coxeter automorphism. If no automorphism is explicitly
specified the twist is realized as ordinary Coxeter twist [19].}
\centering
\vspace{3 mm}
\begin{tabular}{|cr|c|c|rr|rr|r|} \hline
\multicolumn{2}{|c|}{\sf Case }
&\multicolumn{1}{|c|}{\sf Twist}
&\multicolumn{1}{|c|}{\sf Lattice}
&\multicolumn{2}{|c|}{$h^{1,1}$}
&\multicolumn{2}{|c|}{$h^{1,2}$}
&\multicolumn{1}{|c|}{$\chi$} \\[.31 mm] \hline \hline &&&&&&&&\\[-3.5 mm]

1&
$ Z_{3}$&
$(Z_3^{(2)},Z_3^{(2)},Z_3^{(2)})
$ &
$ A_2\times A_2 \times A_2$ &
$ 36  $&$ ( 9  )$&$   -  $&$(-)$&$ 72   $ \\[2 mm]
2&
$ Z_{4}$&
 $(Z_2^{(1)},Z_2^{(1)},Z_4^{(2)}, Z_4^{(2)})
$ &
$ A_1\times A_1\times B_2\times B_2$ &
$  31    $&$ ( 5  )$&$  7 $&$( 1  )$&$ 48   $ \\[2 mm]
3&
$ Z_{4}$&
 $(Z_2^{(1)},Z_4^{(2)}, Z_4^{(3)})
$ &
$A_1\times A_3\times B_2$       &
$ 27     $&$ ( 5  )$&$  3   $&$(  1   )$&$ 48   $ \\[2 mm]
4&
$ Z_{4}$&
 $(Z_4^{(3)}, Z_4^{(3)})
$ &
$A_3\times A_3$&
$ 25     $&$ ( 5  )$&$   1  $&$(   1  )$&$ 48   $ \\[2 mm]
5&
$ Z_{6}$&
 $
(Z_3^{(2)},  Z_6^{(2)} , Z_6^{(2)})
$ &
$A_2\times G_2 \times G_2$ &
$  29    $&$ ( 5  )$&$     5 $&$( -    )$&$  48  $ \\[3 mm]
6&
$ Z_{6}$&
 $(Z_6^{(2)}, Z_3^{(4)})
$ &
$\displaystyle{{G_2 \times A_2\times A_2}\atop{S_1 S_2 S_3 S_4 P_{36}P_{45}}}$&
$    25  $&$ ( 5  )$&$    1 $&$( -    )$&$ 48   $ \\[3 mm]
7&
$ Z^{\prime}_{6}$&
 $
(Z_2^{(1)},Z_2^{(1)},Z_3^{(2)}, Z_6^{(2)})
$ &
$A_1\times A_1\times A_2\times G_2$       &
$  35    $&$ ( 3  )$&$   11  $&$(   1  )$&$  48  $ \\[2 mm]

8&
$ Z^{\prime}_{6}$&
 $(Z_2^{(1)}, Z_3^{(2)}, Z_6^{(3)})
$ &
$A_2\times D_4 $&
$   29   $&$ ( 3  )$&$  5   $&$( 1    )$&$ 48   $ \\[3 mm]
9&
$ Z^{\prime}_{6}$&
 $
(Z_2^{(1)},Z_2^{(1)},Z_3^{(4)})
$ &
$\displaystyle{{A_1\times A_1\times A_2\times A_2}
\atop{S_1 S_2 S_3 S_4 P_{36}P_{45}}}$&
$ 31     $&$ ( 3 )$&$   7  $&$(   1  )$&$ 48   $ \\[3 mm]
10&
$ Z^{\prime}_{6}$&
 $
(Z_2^{(1)}, Z_6^{(5)})
$ &
$ A_1\times A_5$  &
$  25    $&$ (  3 )$&$   1  $&$(  1   )$&$ 48   $ \\[2 mm]
11&
$ Z_{7}$&
 $(Z_7^{(6)})
$ &
$ A_6$       &
$ 24     $&$ ( 3 )$&$   -  $&$(  -   )$&$  48  $ \\[2 mm]
12&
$ Z_{8}$&
 $(Z_4^{(2)},Z_8^{(4)})
$ &
$B_2\times B_4$       &
$   27   $&$ ( 3  )$&$   3  $&$( -    )$&$ 48    $\\[3 mm]
13&
$ Z_{8}$&
 $(Z_8^{(6)})
$ &
$\displaystyle{{A_3\times A_3}
\atop{S_1 S_2 S_3  P_{35}P_{36}P_{45}}}$&
$  24    $&$ ( 3  )$&$  -   $&$(  -   )$&$ 48   $ \\[3 mm]
14&
$ Z^{\prime}_{8}$&
 $(Z_2^{(1)}, Z_2^{(2)}, Z_8^{(4)})
$ &
$B_4\times D_2$ &
$  31    $&$ ( 3  )$&$  7   $&$(  1   )$&$ 48   $ \\[2 mm]
15&
$ Z^{\prime}_{8}$&
 $
(Z_2^{(1)}, Z_8^{(5)})
$ &
$A_1\times D_5$&
$   27   $&$ ( 3  )$&$   3  $&$( 1    )$&$ 48   $ \\[2 mm]
16&
$ Z_{12}$&
 $
(Z_3^{(2)}, Z_{12}^{(4)})
$ &
$A_2\times F_4$&
$ 29     $&$ ( 3  )$&$  5   $&$( -    )$&$  48  $ \\[2 mm]
17&
$ Z_{12}$&
 $
(Z_{12}^{(6)})
$ &
$E_6$&
$   25   $&$ ( 3  )$&$    1 $&$( -    )$&$ 48   $ \\[2 mm]
18&
$ Z^{\prime}_{12}$&
 $
(Z_2^{(1)}, Z_2^{(1)}, Z_{12}^{(4)})
$ &
$D_2\times F_4$ &
$   31   $&$ ( 3  )$&$   7  $&$( 1    )$&$ 48   $ \\[2 mm] \hline
\end{tabular}
\label{orbtab}
\end{table}

\begin{figure}
\setlength{\unitlength}{ 4 mm}
\caption{Schematic configuration of the orbifold singularities.
Fixed point singularites are depicted by dots,
fixed torus singularities by lines. We indicate
the maximal order of the group element under
which the sets stay fix in parantheses.
The numbers on the sets indicates their multiplicity
on the torus.}
\vskip 3 mm

\begin{picture}(12.5,10)(0,0)
\put(0,6.5){1.) $Z_3$:}
\put(5.3,6.5){\circle*{0.3}}
\put(5.8,6.5){$^{(3)}$}

\put(0,  0.5){\circle*{0.3}}
\put(0.8,0.5){\circle*{0.3}}
\put(1.6,0.5){\circle*{0.3}}
\put(2.4,0.5){\circle*{0.3}}
\put(3.2,0.5){\circle*{0.3}}
\put(4  ,0.5){\circle*{0.3}}
\put(4.8,0.5){\circle*{0.3}}
\put(5.6,0.5){\circle*{0.3}}
\put(6.4,0.5){\circle*{0.3}}

\put(0  ,1.5){\circle*{0.3}}
\put(0.8,1.5){\circle*{0.3}}
\put(1.6,1.5){\circle*{0.3}}
\put(2.4,1.5){\circle*{0.3}}
\put(3.2,1.5){\circle*{0.3}}
\put(4  ,1.5){\circle*{0.3}}
\put(4.8,1.5){\circle*{0.3}}
\put(5.6,1.5){\circle*{0.3}}
\put(6.4,1.5){\circle*{0.3}}

\put(0  ,2.5){\circle*{0.3}}
\put(0.8,2.5){\circle*{0.3}}
\put(1.6,2.5){\circle*{0.3}}
\put(2.4,2.5){\circle*{0.3}}
\put(3.2,2.5){\circle*{0.3}}
\put(4,  2.5){\circle*{0.3}}
\put(4.8,2.5){\circle*{0.3}}
\put(5.6,2.5){\circle*{0.3}}
\put(6.4,2.5){\circle*{0.3}}

\end{picture}
\begin{picture}(12.5,10)(0,0)
\put(0,6.5){2.) $Z_4$:}
\put(5.3,6.5){\circle*{0.3}}
\put(5.8,6.5){$^{(4)}$}
\put(7.5,6.5){\line(1,0){.5}}
\put(8.1,6.5){$^{(2)}$}

\put(0,0.5){\line(1,0){4}}
\put(0.8,0.5){\circle*{0.3}}
\put(1.6,0.5){\circle*{0.3}}
\put(2.4,0.5){\circle*{0.3}}
\put(3.2,0.5){\circle*{0.3}}

\put(0  ,1){\line(1,0){4}}
\put(0.8,1){\circle*{0.3}}
\put(1.6,1){\circle*{0.3}}
\put(2.4,1){\circle*{0.3}}
\put(3.2,1){\circle*{0.3}}

\put(0,  1.5){\line(1,0){4}}
\put(0.8,1.5){\circle*{0.3}}
\put(1.6,1.5){\circle*{0.3}}
\put(2.4,1.5){\circle*{0.3}}
\put(3.2,1.5){\circle*{0.3}}

\put(0,  2){\line(1,0){4}}
\put(0.8,2){\circle*{0.3}}
\put(1.6,2){\circle*{0.3}}
\put(2.4,2){\circle*{0.3}}
\put(3.2,2){\circle*{0.3}}

\put(5,0.5){\line(1,0){3}}
\put(5,1  ){\line(1,0){3}}
\put(5,1.5){\line(1,0){3}}
\put(5,2  ){\line(1,0){3}}
\put(5,2.5  ){\line(1,0){3}}
\put(5,3 ){\line(1,0){3}}
\put(8.2,1.5){$ \Biggl\}$ {\scriptsize 2}}

\end{picture}
\begin{picture}(12.5,10)(0,0)
\put(0,6.5){3.) $Z_4$: }
\put(5.3,6.5){\circle*{0.3}}
\put(5.8,6.5){$^{(4)}$}
\put(7.5,6.5){\line(1,0){.5}}
\put(8.1,6.5){$^{(2)}$}

\put(0,0.5){\line(1,0){4}}
\put(0.8,0.5){\circle*{0.3}}
\put(1.6,0.5){\circle*{0.3}}
\put(2.4,0.5){\circle*{0.3}}
\put(3.2,0.5){\circle*{0.3}}

\put(0  ,1){\line(1,0){4}}
\put(0.8,1){\circle*{0.3}}
\put(1.6,1){\circle*{0.3}}
\put(2.4,1){\circle*{0.3}}
\put(3.2,1){\circle*{0.3}}

\put(0,  1.5){\line(1,0){4}}
\put(0.8,1.5){\circle*{0.3}}
\put(1.6,1.5){\circle*{0.3}}
\put(2.4,1.5){\circle*{0.3}}
\put(3.2,1.5){\circle*{0.3}}

\put(0,  2){\line(1,0){4}}
\put(0.8,2){\circle*{0.3}}
\put(1.6,2){\circle*{0.3}}
\put(2.4,2){\circle*{0.3}}
\put(3.2,2){\circle*{0.3}}

\put(5,0.5){\line(1,0){3}}
\put(5,1  ){\line(1,0){3}}
\put(8.2,0.5){$ \}$ {\scriptsize 2} }

\end{picture}
\begin{picture}(12.5,11)(0,0)
\put(0,6.5){4.) $Z_4$: }
\put(5.3,6.5){\circle*{0.3}}
\put(5.8,6.5){$^{(4)}$}
\put(7.5,6.5){\line(1,0){.5}}
\put(8.1,6.5){$^{(2)}$}

\put(0,0.5){\line(1,0){4}}
\put(0.8,0.5){\circle*{0.3}}
\put(1.6,0.5){\circle*{0.3}}
\put(2.4,0.5){\circle*{0.3}}
\put(3.2,0.5){\circle*{0.3}}

\put(0  ,1){\line(1,0){4}}
\put(0.8,1){\circle*{0.3}}
\put(1.6,1){\circle*{0.3}}
\put(2.4,1){\circle*{0.3}}
\put(3.2,1){\circle*{0.3}}

\put(0,  1.5){\line(1,0){4}}
\put(0.8,1.5){\circle*{0.3}}
\put(1.6,1.5){\circle*{0.3}}
\put(2.4,1.5){\circle*{0.3}}
\put(3.2,1.5){\circle*{0.3}}

\put(0,  2){\line(1,0){4}}
\put(0.8,2){\circle*{0.3}}
\put(1.6,2){\circle*{0.3}}
\put(2.4,2){\circle*{0.3}}
\put(3.2,2){\circle*{0.3}}


\end{picture}
\begin{picture}(12.5,11)(0,0)
\put(0,6.5){5.) $Z_6$:}
\put(5.3,6.5){\circle*{0.3}}
\put(5.8,6.5){$^{(6)}$}
\put(7.5,6.5){\circle{0.3}}
\put(7.8,6.5){$^{(3)}$}
\put(5.3,5.5){\line(1,0){.5}}
\put(5.9,5.5){$^{(2)}$}
\put(0   ,0.5){\line(1,0){4}}
\put(1.25,0.5){\circle*{0.3}}
\put(2   ,0.5){\circle*{0.3}}
\put(2.75,0.5){\circle*{0.3}}

\put(1.25,2  ){\circle{0.3}}
\put(2   ,2  ){\circle{0.3}}
\put(2.75,2  ){\circle{0.3}}

\put(1.25,2.5  ){\circle{0.3}}
\put(2   ,2.5  ){\circle{0.3}}
\put(2.75,2.5  ){\circle{0.3}}

\put(1.25,3  ){\circle{0.3}}
\put(2   ,3  ){\circle{0.3}}
\put(2.75,3  ){\circle{0.3}}

\put(1.25,3.5  ){\circle{0.3}}
\put(2   ,3.5  ){\circle{0.3}}
\put(2.75,3.5  ){\circle{0.3}}
\put(1.15,3.1){$\overbrace{\phantom{--,}}^2$}

\put(5,0.5){\line(1,0){3}}
\put(5,1  ){\line(1,0){3}}
\put(5,1.5){\line(1,0){3}}
\put(5,2  ){\line(1,0){3}}
\put(5,2.5  ){\line(1,0){3}}
\put(8.2,1.3){$ \biggl\}$ {\scriptsize 3}}

\end{picture}
\begin{picture}(12.5,11)(0,0)
\put(0,6.5){6.) $Z_6$:}
\put(5.3,6.5){\circle*{0.3}}
\put(5.8,6.5){$^{(6)}$}
\put(7.5,6.5){\circle{0.3}}
\put(7.8,6.5){$^{(3)}$}
\put(5.3,5.5){\line(1,0){.5}}
\put(5.9,5.5){$^{(2)}$}
\put(0   ,0.5){\line(1,0){4}}
\put(1.25,0.5){\circle*{0.3}}
\put(2   ,0.5){\circle*{0.3}}
\put(2.75,0.5){\circle*{0.3}}

\put(1.25,2  ){\circle{0.3}}
\put(2   ,2  ){\circle{0.3}}
\put(2.75,2  ){\circle{0.3}}

\put(1.25,2.5  ){\circle{0.3}}
\put(2   ,2.5  ){\circle{0.3}}
\put(2.75,2.5  ){\circle{0.3}}

\put(1.25,3  ){\circle{0.3}}
\put(2   ,3  ){\circle{0.3}}
\put(2.75,3  ){\circle{0.3}}

\put(1.25,3.5  ){\circle{0.3}}
\put(2   ,3.5  ){\circle{0.3}}
\put(2.75,3.5  ){\circle{0.3}}
\put(1.15,3.1){$\overbrace{\phantom{--,}}^2$}
\put(5,0.5){\line(1,0){3}}
\put(8.2,0.25){ {\scriptsize 3}}

\end{picture}
\begin{picture}(12.5,14)(0,0)
\put(0,9.5){7.) $Z_6'$:}
\put(5.3,9.5){\circle*{0.3}}
\put(5.8,9.5){$^{(6)}$}
\put(7.5,9.5){\line(0,1){.5}}
\put(7.8,9.5){$^{(3)}$}
\put(5.3,8.5){\line(1,0){.5}}
\put(5.9,8.5){$^{(2)}$}
\put(0,1.5){\line(1,0){4}}
\put(1.25,1.5){\circle*{0.3}}
\put(2,1.5){\circle*{0.3}}
\put(2.75,1.5){\circle*{0.3}}
\put(0,2  ){\line(1,0){4}}
\put(1.25,2  ){\circle*{0.3}}
\put(2,2  ){\circle*{0.3}}
\put(2.75,2  ){\circle*{0.3}}
\put(0,2.5){\line(1,0){4}}
\put(1.25,2.5){\circle*{0.3}}
\put(2,2.5){\circle*{0.3}}
\put(2.75,2.5){\circle*{0.3}}
\put(0,3  ){\line(1,0){4}}
\put(1.25,3  ){\circle*{0.3}}
\put(2,3  ){\circle*{0.3}}
\put(2.75,3  ){\circle*{0.3}}
\put(1.25,0.5){\line(0,1){3.5}}
\put(2,0.5){\line(0,1){3.5}}
\put(2.75,0.5){\line(0,1){3.5}}
\put(1.24,6.5){$\overbrace{\phantom{--}}^2$}
\put(1.25,5){\line(0,1){2}}
\put(2,5){\line(0,1){2}}
\put(2.75,5){\line(0,1){2}}
\put(5,1.5){\line(1,0){3}}
\put(5,2  ){\line(1,0){3}}
\put(5,2.5){\line(1,0){3}}
\put(5,3  ){\line(1,0){3}}
\put(8.2,2){$ \Bigl\}$ {\scriptsize 3} }
\end{picture}
\begin{picture}(12.5,14)(0,0)
\put(0,9.5){8.) $Z_6'$:}
\put(5.4,9.5){\circle*{0.3}}
\put(5.9,9.5){$^{(6)}$}
\put(7.6,9.5){\line(0,1){.5}}
\put(7.9,9.5){$^{(3)}$}
\put(5.4,8.5){\line(1,0){.5}}
\put(6,  8.5){$^{(2)}$}
\put(0,1.5){\line(1,0){4}}
\put(1.25,1.5){\circle*{0.3}}
\put(2,1.5){\circle*{0.3}}
\put(2.75,1.5){\circle*{0.3}}
\put(0,2  ){\line(1,0){4}}
\put(1.25,2  ){\circle*{0.3}}
\put(2,2  ){\circle*{0.3}}
\put(2.75,2  ){\circle*{0.3}}
\put(0,2.5){\line(1,0){4}}
\put(1.25,2.5){\circle*{0.3}}
\put(2,2.5){\circle*{0.3}}
\put(2.75,2.5){\circle*{0.3}}
\put(0,3  ){\line(1,0){4}}
\put(1.25,3  ){\circle*{0.3}}
\put(2,3  ){\circle*{0.3}}
\put(2.75,3  ){\circle*{0.3}}
\put(1.25,0.5){\line(0,1){3.5}}
\put(2,0.5){\line(0,1){3.5}}
\put(2.75,0.5){\line(0,1){3.5}}
\put(5,1.5){\line(1,0){3}}
\put(5,2  ){\line(1,0){3}}
\put(5,2.5){\line(1,0){3}}
\put(5,3  ){\line(1,0){3}}
\put(8.2,2){$ \Bigl\}$ {\scriptsize 3}}
\end{picture}
\begin{picture}(12.5,14)(0,0)
\put(0,9.5){9.) $Z_6'$:}
\put(5,9.5){\circle*{0.3}}
\put(5.5,9.5){$^{(6)}$}
\put(7.5,9.5){\line(0,1){.5}}
\put(7.8,9.5){$^{(3)}$}
\put(5,8.5){\line(1,0){.5}}
\put(5.6,8.5){$^{(2)}$}
\put(0,1.5){\line(1,0){4}}
\put(1.25,1.5){\circle*{0.3}}
\put(2,1.5){\circle*{0.3}}
\put(2.75,1.5){\circle*{0.3}}
\put(0,2  ){\line(1,0){4}}
\put(1.25,2  ){\circle*{0.3}}
\put(2,2  ){\circle*{0.3}}
\put(2.75,2  ){\circle*{0.3}}
\put(0,2.5){\line(1,0){4}}
\put(1.25,2.5){\circle*{0.3}}
\put(2,2.5){\circle*{0.3}}
\put(2.75,2.5){\circle*{0.3}}
\put(0,3  ){\line(1,0){4}}
\put(1.25,3  ){\circle*{0.3}}
\put(2,3  ){\circle*{0.3}}
\put(2.75,3  ){\circle*{0.3}}
\put(1.25,0.5){\line(0,1){3.5}}
\put(2,0.5){\line(0,1){3.5}}
\put(2.75,0.5){\line(0,1){3.5}}

\put(1.25,5){\line(0,1){2}}
\put(2,5){\line(0,1){2}}
\put(2.75,5){\line(0,1){2}}
\put(1.15,6.5){$\overbrace{\phantom{--}}^2$}
\end{picture}
\begin{picture}(12.5,11)(0,0)
\put(0,6.5){10.) $Z_6'$:}
\put(5,6.5){\circle*{0.3}}
\put(5.5,6.5){$^{(6)}$}
\put(7.5,6.5){\line(0,1){.5}}
\put(7.8,6.5){$^{(3)}$}
\put(5,5.5){\line(1,0){.5}}
\put(5.6,5.5){$^{(2)}$}
\put(0,1.5){\line(1,0){4}}
\put(1.25,1.5){\circle*{0.3}}
\put(2,1.5){\circle*{0.3}}
\put(2.75,1.5){\circle*{0.3}}
\put(0,2  ){\line(1,0){4}}
\put(1.25,2  ){\circle*{0.3}}
\put(2,2  ){\circle*{0.3}}
\put(2.75,2  ){\circle*{0.3}}
\put(0,2.5){\line(1,0){4}}
\put(1.25,2.5){\circle*{0.3}}
\put(2,2.5){\circle*{0.3}}
\put(2.75,2.5){\circle*{0.3}}
\put(0,3  ){\line(1,0){4}}
\put(1.25,3  ){\circle*{0.3}}
\put(2,3  ){\circle*{0.3}}
\put(2.75,3  ){\circle*{0.3}}
\put(1.25,0.5){\line(0,1){3.5}}
\put(2,0.5){\line(0,1){3.5}}
\put(2.75,0.5){\line(0,1){3.5}}
\end{picture}
\begin{picture}(12.5,11)(0,0)
\put(0,6.5){11.) $Z_7$:}
\put(5.3,6.5){\circle*{0.3}}
\put(5.8,6.5){$^{(7)}$}

\put(2,  4.3){\circle*{0.3}}
\put(3.56,3.55){\circle*{0.3}}
\put(0.44,3.55){\circle*{0.3}}
\put(3.95,1.75){\circle*{0.3}}
\put(0.05,1.75){\circle*{0.3}}
\put(1.13,0.5){\circle*{0.3}}
\put(2.87,0.5){\circle*{0.3}}
\end{picture}
\begin{picture}(12.5,12)(0,0)
\put(0,6.5){12.) $Z_8$:}
\put(5.3,6.5){\circle*{0.3}}
\put(5.8,6.5){$^{(8)}$}
\put(7.5,6.5){\circle{0.3}}
\put(8  ,6.5){$^{(4)}$}
\put(5.3,5.5){\line(1,0){.5}}
\put(5.9,5.5){$^{(2)}$}

\put(0,0.5){\line(1,0){4}}
\put(0.8,0.5){\circle*{0.3}}
\put(1.6,0.5){\circle*{0.3}}
\put(2.8,0.5){\circle{0.3}}
\put(2.6,-0.2){\scriptsize 2}
\put(0  ,1){\line(1,0){4}}
\put(0.8,1){\circle*{0.3}}
\put(1.6,1){\circle*{0.3}}
\put(2.8,1){\circle{0.3}}
\put(2.6,1.3){\scriptsize 2}


\put(0,  2.5){\line(1,0){4}}
\put(0.8,2.5){\circle{0.3}}
\put(1.6,2.5){\circle{0.3}}
\put(2.4,2.5){\circle{0.3}}
\put(3.2,2.5){\circle{0.3}}
\put(4.2,2.25){ {\scriptsize 2}}

\put(5,0.5){\line(1,0){3}}
\put(5,1  ){\line(1,0){3}}
\put(5,1.5){\line(1,0){3}}
\put(8.2,.75){$ \big\} \,$ {\scriptsize 4}}

\end{picture}
\end{figure}
\newpage
\begin{figure}
\setlength{\unitlength}{ 4 mm}
\begin{picture}(12.5,12)(0,0)
\put(0,8.5){13.) $Z_8$:}
\put(5.3,8.5){\circle*{0.3}}
\put(5.8,8.5){$^{(8)}$}
\put(7.5,8.5){\circle{0.3}}
\put(8  ,8.5){$^{(4)}$}
\put(5.3,7.5){\line(1,0){.5}}
\put(5.9,7.5){$^{(2)}$}

\put(0,0.5){\line(1,0){4}}
\put(0.8,0.5){\circle*{0.3}}
\put(1.6,0.5){\circle*{0.3}}
\put(2.8,0.5){\circle{0.3}}
\put(2.6,-0.2){\scriptsize 2}
\put(0  ,1){\line(1,0){4}}
\put(0.8,1){\circle*{0.3}}
\put(1.6,1){\circle*{0.3}}
\put(2.8,1){\circle{0.3}}
\put(2.6,1.3){\scriptsize 2}

\put(0,  2.5){\line(1,0){4}}
\put(0.8,2.5){\circle{0.3}}
\put(1.6,2.5){\circle{0.3}}
\put(2.4,2.5){\circle{0.3}}
\put(3.2,2.5){\circle{0.3}}
\put(4.2,2.25){ {\scriptsize 2}}


\end{picture}
\begin{picture}(12.5,12)(0,0)
\put(0,8.5){14.) $Z_8'$:}
\put(5.3,8.5){\circle*{0.3}}
\put(5.8,8.5){$^{(8)}$}
\put(7.5,8.5){\line(1,0){0.5}}
\put(8.1,8.5){$^{(4)}$}
\put(5.3,7.5){\line(0,1){.5}}
\put(5.6,7.5){$^{(2)}$}

\put(0,0.5){\line(1,0){4}}
\put(0.8,0.5){\circle*{0.3}}
\put(1.6,0.5){\circle*{0.3}}
\put(2.4,0.5){\circle*{0.3}}
\put(3.2,0.5){\circle*{0.3}}

\put(0,  1){\line(1,0){4}}
\put(0.8,1){\circle*{0.3}}
\put(1.6,1){\circle*{0.3}}
\put(2.4,1){\circle*{0.3}}
\put(3.2,1){\circle*{0.3}}

\put(5,0.5){\line(1,0){3}}
\put(8.2,0.25){\scriptsize 2}

\put(1.25,3){\line(0,1){2}}
\put(2,3){\line(0,1){2}}
\put(2.75,3){\line(0,1){2}}
\put(1.15,4.7){$\overbrace{\phantom{--}}^4$}
\end{picture}
\begin{picture}(12.5,12)(0,0)
\put(0,8.5){15.) $Z_8'$:}
\put(5.3,8.5){\circle*{0.3}}
\put(5.8,8.5){$^{(8)}$}
\put(7.5,8.5){\line(1,0){0.5}}
\put(8.1,8.5){$^{(4)}$}
\put(5.3,7.5){\line(0,1){.5}}
\put(5.6,7.5){$^{(2)}$}

\put(0,0.5){\line(1,0){4}}
\put(0.8,0.5){\circle*{0.3}}
\put(1.6,0.5){\circle*{0.3}}
\put(2.4,0.5){\circle*{0.3}}
\put(3.2,0.5){\circle*{0.3}}

\put(0,  1){\line(1,0){4}}
\put(0.8,1){\circle*{0.3}}
\put(1.6,1){\circle*{0.3}}
\put(2.4,1){\circle*{0.3}}
\put(3.2,1){\circle*{0.3}}

\put(1.5,3){\line(0,1){2}}
\put(1.3,5.2){{\scriptsize 4}}
\put(2.5,3){\line(0,1){2}}
\put(2.3,5.2){{\scriptsize 2}}

\end{picture}
\begin{picture}(12.5,14)(0,0)
\put(0,8.5){16.) $Z_{12}$:}
\put(5.3,8.5){\circle*{0.3}}
\put(5.8,8.5){$^{(12)}$}
\put(7.5,8.5){\line(1,0){.5}}
\put(8.1,8.5){$^{(4)}$}
\put(5.3,7.5){\circle{0.3}}
\put(5.9,7.5){$^{(3)}$}
\put(7.3,7.5){\line(0,1){.5}}
\put(7.6,7.5){$^{(2)}$}

\put(0   ,0.5){\line(1,0){4}}
\put(1.25,0.5){\circle*{0.3}}
\put(2   ,0.5){\circle*{0.3}}
\put(2.75,0.5){\circle*{0.3}}

\put(1.25,2  ){\circle{0.3}}
\put(2   ,2  ){\circle{0.3}}
\put(2.75,2  ){\circle{0.3}}

\put(1.25,2.5  ){\circle{0.3}}
\put(2   ,2.5  ){\circle{0.3}}
\put(2.75,2.5  ){\circle{0.3}}
\put(1.15,2.2){$\overbrace{\phantom{--,}}^4$}

\put(5,0.5){\line(1,0){3}}
\put(8.2,0.25){\scriptsize 3}
\put(6,2){\line(0,1){2}}
\put(7,2){\line(0,1){2}}
\put(5.8,4.2){\scriptsize 6}
\put(6.8,4.2){\scriptsize 6}
\end{picture}
\begin{picture}(12.5,14)(0,0)
\put(0,8.5){17.) $Z_{12}$:}
\put(5.3,8.5){\circle*{0.3}}
\put(5.8,8.5){$^{(12)}$}
\put(7.5,8.5){\line(1,0){.5}}
\put(8.1,8.5){$^{(4)}$}
\put(5.3,7.5){\circle{0.3}}
\put(5.9,7.5){$^{(3)}$}
\put(7.3,7.5){\line(0,1){.5}}
\put(7.6,7.5){$^{(2)}$}

\put(0   ,0.5){\line(1,0){4}}
\put(1.25,0.5){\circle*{0.3}}
\put(2   ,0.5){\circle*{0.3}}
\put(2.75,0.5){\circle*{0.3}}

\put(1.25,2  ){\circle{0.3}}
\put(2   ,2  ){\circle{0.3}}
\put(2.75,2  ){\circle{0.3}}

\put(1.25,2.5  ){\circle{0.3}}
\put(2   ,2.5  ){\circle{0.3}}
\put(2.75,2.5  ){\circle{0.3}}
\put(1.15,2.2){$\overbrace{\phantom{--,}}^4$}

\put(6.5,2){\line(0,1){2}}
\put(6.3,4.2){\scriptsize 3}
\end{picture}
\begin{picture}(12.5,12)(0,0)
\put(0,8.5){18.) $Z_{12}'$:}
\put(5.3,8.5){\circle*{0.3}}
\put(5.8,8.5){$^{(12)}$}
\put(7.5,8.5){\line(1,0){.5}}
\put(8.1,8.5){$^{(6)}$}
\put(5.3,7.5){\circle{0.3}}
\put(5.9,7.5){$^{(4)}$}
\put(7.5,7.5){\line(1,1){0.8}}
\put(8,  7.5){$^{(3)}$}
\put(5.3,6.5){\line(0,1){.5}}
\put(5.6,6.5){$^{(2)}$}

\put(0   ,0.5){\line(1,0){4}}
\put(0.8,0.5){\circle*{0.3}}
\put(1.6  ,0.5){\circle*{0.3}}
\put(2.4,0.5){\circle*{0.3}}
\put(3.2,0.5){\circle*{0.3}}

\put(1.25,2){\line(0,1){4}}
\put(2,2  ){\line(0,1){4}}
\put(1.05 ,6.2){\scriptsize 6}
\put(1.8 ,6.2){\scriptsize 6}
\put(2.75,2){\line(0,1){4}}
\put(2.55,6.2){\scriptsize 3}
\put(2.75,2.8 ){\circle{0.3}}
\put(2.75,3.6 ){\circle{0.3}}
\put(2.75,4.4 ){\circle{0.3}}
\put(2.75,5.2 ){\circle{0.3}}

\put(5.5,0.5){\line(1,1){3}}
\put(5,1  ){\line(1,1){3}}
\put(8.4   ,3.6){\scriptsize 4}
\put(8 ,4.2){\scriptsize 4}

\end{picture}
\end{figure}

\clearpage

\centerline{Desingularisation of the nine types of $Z_N$ singular points:}
\vskip 180 true pt
\centerline{
$Z_3:\, {1\over 3} (1,1,1)$
\qquad\qquad\qquad\qquad
$Z_4:\, {1\over 4} (1,1,2)$
\qquad\qquad\qquad\qquad
$Z_6:\, {1\over 3} (1,1,4)$ }
\vskip 192 true pt
\centerline{
$Z_6':\, {1\over 6} (1,2,3)$
\qquad\qquad\qquad\qquad
$Z_7:\, {1\over 7} (1,2,4)$
\qquad\qquad\qquad\qquad
$Z_8:\, {1\over 8} (1,5,2)$ }
\vskip 192 true pt
\centerline{
$Z_8':\, {1\over 8} (1,3,4)$
\qquad\qquad\qquad\qquad
$Z_{12}:\, {1\over 12} (1,7,4)$
\qquad\qquad\qquad\qquad
$Z_{12}':\, {1\over 12} (1,5,6)$ }

\vskip -574  true pt

\includegraphics{3:111.ps}
\includegraphics{4:112.ps}
\includegraphics{6:114.ps}
\includegraphics{6:123.ps}
\includegraphics{7:124.ps}
\includegraphics{8:125.ps}
\includegraphics{8:134.ps}
\includegraphics{12:147.ps}
\includegraphics{12:156.ps}
\vskip 16.5 true cm

\end{document}